\documentclass[12pt]{iopart}

\usepackage{xcolor,hyperref,acronym}
\usepackage{ifthen,comment}

\usepackage[square,numbers,sort&compress]{natbib}

\usepackage{iopams}
\expandafter\let\csname equation*\endcsname\relax
\expandafter\let\csname endequation*\endcsname\relax
\usepackage{amsmath}
\usepackage{acronym}

\begin{document}

\title[White Paper]{White Paper and Roadmap for Quantum Gravity Phenomenology in the Multi-Messenger Era}

\author{R~Alves~Batista$^{1,98}$,
G~Amelino-Camelia$^{2,3}$,
D~Boncioli$^{4,5}$,
J~M~Carmona$^{6}$,
A~di~Matteo$^{7}$,
G~Gubitosi$^{2,3}$,
I~Lobo$^{8,9}$,
N~E~Mavromatos$^{10,11}$,
C~Pfeifer$^{12}$,
D~Rubiera-Garcia$^{13}$
% ...
}
\frenchspacing 

\address{$^{1}$ Instituto de F\'isica Te\'orica UAM-CSIC, Universidad Aut\'onoma de Madrid, C/ Nicol\'as Cabrera 13-15, 28049 Madrid, Spain}
\address{$^{2}$ Dipartimento di Fisica “Ettore Pancini”, Universit\'a di Napoli “Federico II”, Complesso Univ. Monte S. Angelo, I-80126 Napoli, Italy}
\address{$^{3}$ Istituto Nazionale di Fisica Nucleare, Sezione di Napoli, Complesso Univ. Monte S. Angelo, I-80126 Napoli, Italy}
\address{$^{4}$ Università degli Studi dell'Aquila, Dipartimento di Scienze Fisiche e Chimiche, via Vetoio, 67100 L'Aquila, Italy}
\address{$^{5}$ Istituto Nazionale di Fisica Nucleare, Laboratori Nazionali del Gran Sasso, via Acitelli, 67100 Assergi (AQ), Italy}
\address{$^{6}$ Centro de Astropartículas y Física de Altas Energías (CAPA), Departamento de Física Teórica, Universidad de Zaragoza, C/ Pedro Cerbuna 12, 50009 Zaragoza, Spain}
\address{$^{7}$ Istituto Nazionale di Fisica Nucleare, Sezione di Torino, Via Pietro Giuria 1, 10125 Torino, Italy}
\address{$^{8}$ Federal University of Para\'iba, Rodovia BR 079 - km 12, 58397-000 Areia-PB, Brazil}
\address{$^{9}$ Federal University of Lavras, Caixa Postal 3037, 37200-900 Lavras-MG, Brazil}
\address{$^{10}$ National Technical University of Athens, School of Applied Mathematical and Physical Sciences, Department of Physics, 9 Iroon Polytechniou Str., Zografou Campus 157 80, Athens, Greece}
\address{$^{11}$ King's College London, Department of Physics, Strand, London WC2R 2LS, UK}
\address{$^{12}$ Center of Applied Space Technology and Microgravity (ZARM), University of Bremen, Am Fallturm 2 28359 Bremen ,Germany}
\address{$^{13}$ Departamento de F\'isica Te\'orica and IPARCOS, Universidad Complutense de Madrid , E-28040 Madrid, Spain}
% ...
\address{$^{98}$ Institut d’Astrophysique de Paris,CNRS, Sorbonne Université, Paris 75014, France}
% ...
\nonfrenchspacing
\eads{\mailto{christian.pfeifer@zarm.uni-bremen.de}, \mailto{qgmm.cost@gmail.com}}

\begin{abstract}
The unification of quantum mechanics and general relativity has long been elusive.  Only recently have empirical predictions of various possible theories of quantum gravity been put to test, where a clear signal of quantum properties of gravity is still missing.  The dawn of multi-messenger high-energy astrophysics has been tremendously beneficial, as it allows us to study particles with much higher energies and travelling much longer distances than possible in terrestrial experiments, but more progress is needed on several fronts. 

A thorough appraisal of current strategies and experimental frameworks, regarding quantum gravity phenomenology, is provided here. Our aim is twofold: a description of tentative multimessenger explorations, plus a focus on future detection experiments. 

As the outlook of the network of researchers that formed through the COST Action CA18108 ``Quantum gravity phenomenology in the multi-messenger approach (QG-MM)'', in this work we give an overview of the desiderata that future theoretical frameworks, observational facilities, and data-sharing policies should satisfy in order to advance the cause of quantum gravity phenomenology.
\end{abstract}

%
\vspace{2pc}

\submitto{\CQG}
%
%
%
\tableofcontents

\section{Introduction}
Is gravity quantized? If not, why does gravity have fundamentally different properties from the other three fundamental forces of nature? If yes, how does a fundamental self-consistent theory of \ac{QG} look and where does the quantum nature of gravity manifest itself in observations? These are among the major open questions in fundamental physics.

There are many theoretical approaches aiming at finding a self-consistent theory of \ac{QG}, such as canonical quantum gravity \cite{Kiefer:2004xyv}, string theory \cite{Polchinski:1998rq,Polchinski:1998rr}, loop quantum gravity \cite{Ashtekar:2021kfp}, non-commutative geometry \cite{Connes:1996gi}, asymptotic safety \cite{Eichhorn:2018yfc}, causal dynamical triangulation \cite{Loll:2019rdj}, or causal sets \cite{Surya:2019ndm}, all focusing on different aspects of how the theory should look like, inspired by what we know about the mathematical structures of quantum theory. The path from these fundamental approaches to quantitative predictions of observables often takes huge computational efforts and requires simplifying assumptions and approximations.

What is missing is a clear signal  of quantum properties of gravity, the reasons for which might be that we have not yet constructed the right dedicated experiment to search for \ac{QG} effects, that the current instruments do not yet have the sensitivity needed to detect Planck scale effects, or that some effects have been overlooked in available data.

To bridge this gap between fundamental theories of \ac{QG} and observations, a bottom-up approach emerged: \ac{QG} phenomenology \cite{Addazi:2021xuf}. It is a framework that employs phenomenological models inspired by the results from the different fundamental approaches and by physical reasoning about \ac{QG}, to identify effects which can be searched for in experiments. How and which of these effects emerge from the theory, depends on the theoretical model under consideration, for example: semi-classical field equations; the coupling between local tabletop low-energy quantum systems and gravity; modified gravity with additional scalar or vector fields or modified gravity with additional geometric fields such as torsion, non-metricity or non-linear connections; non-minimal couplings between gravity and matter; deviations from Lorentz invariance as in the standard model extension (SME) or in doubly special relativity (DSR); and many more \cite{Carlip:2008zf,Carney:2018ofe,Pfeifer:2019zhc,Arzano:2021scz,Addazi:2021xuf,ModQuantGrav,Kostelecky:2020hbb,Abdalla:2023gmc,Lehnert2023}. These models help to devise the experimental setups required for searching for specific signatures, and the types of effects being sought. At the same time, evidence or bounds on any given effect are serving as guides for the construction of fundamental theories compatible with these phenomenological results.

In the last years, the COST Action CA18108 ``Quantum Gravity Phenomenology in the Multi-Messenger Approach''\footnote{see for more information, \url{https://qg-mm.unizar.es/}} gathered physicists from different backgrounds in the theory of \ac{QG} and multi-messenger astronomy, searching for possible imprints of \ac{QG} in the combined observational data of different cosmic messengers, namely gamma rays, neutrinos, \acp{UHECR} and gravitational waves. It became clear that one major obstacle in finding them is that often the searches for \ac{QG} are not the primary goal of existing experimental setups and observatories, which were originally designed for another purpose. Further difficulties to proceed arise from a restricted data availability and a language and interaction barrier between researchers of different disciplines (in particular experiment and theory). 

In this white paper the \ac{QG} phenomenology community summarizes and highlights paths to advance and improve the search for \ac{QG} effects. The overall purpose of this white paper is to outline the future phenomenological and experimental research of \ac{QG} with astrophysical data. 

We address some of the most promising observational windows to scales where we expect \ac{QG} effects to set in, where we focus on selected effects caused by violations, deviations or modifications from Lorentz invariance, since these leave observables traces well below the Planck scale and are, if present, in reach of being either detected or falsified by multi-messenger astronomy observations in the near future. Moreover, we identify the necessary conditions for realistically boosting these searches with present and future facilities. Even more importantly, we aspire to stimulate developments of novel methods for searches for signatures of \ac{QG}. Given that at this point there are no facilities designed for this specific purpose, and experimental tests are usually performed as side projects, we particularly hope to motivate a development of dedicated experiments, and that this white paper can be used as a road map in such endeavours. Surely, this white paper cannot cover all the aspects of the search for \ac{QG} so we focus on selected promising prospects of \ac{QG} searches in multi-messenger astronomy.

The structure of the article is the following. First, in \sref{sec:WhatEffects}, we present a discussion on which kind of effects emerge from different approaches to \ac{QG} that are most promising to look for. In the following \sref{sec:expReq}, we identify the experimental requirements needed to unambiguously identify the presence or absence of these effects in observational data. Here, it is important to find a way to distinguish between \ac{QG} and competing effects. This section serves as preparation to propose dedicated new experiments in \sref{sec:proposals} in order to search for \ac{QG}. In \sref{sec:data}, it is argued how to handle data availability is important for good scientific progress. Finally, in \sref{sec:final} we summarize our discussion and conclude.

\section{What effects are we looking for?}
\label{sec:WhatEffects}

There are numerous effects which emerge from theoretical models of quantum gravity. We discuss the general philosophy about quantum spacetimes in \sref{S:stf}, before we highlight time delay effects, caused by the propagation of particles and fields on quantum spacetime in \sref{ssec:TD}, and interaction effects caused by the interaction of particles and fields on quantum spacetime in \sref{Sec:InteractionAnomalies}.

\subsection{Quantum fluctuating spacetime (``foam'')}
\label{S:stf}

Despite the absence of a generally accepted framework of \ac{QG}, various candidate models seem to converge on the idea that, due to fundamental quantum uncertainties, the microstructure of spacetime should be viewed as a dynamical entity fluctuating over distances of the order of Planck length~$\ell_\text{P}$ and time scales of the order of Planck time~$t_\text{P}$. This idea, originally introduced by Wheeler \cite{Wheeler:1955zz}, further developed  by Hawking \cite{Hawking:1978pog}, and popularized by Wheeler again \cite{10.2307/986184}, implies that, if such fluctuations are large enough to induce non-trivial deformations of the classical, smooth spacetime, the latter would develop a ``foamy'' structure at the ultra-microscopic level, with all manners of geometrically and topologically nontrivial structures being formed (e.g.\ via quantum tunnelling \cite{Garfinkle:1990eq}), evolving, interacting and lasting only a few Planck times.  Closely related to this \textit{spacetime foam} is the idea of \textit{emergent gravity}, by which the classical continuous gravitational field is not fundamental, but instead emerges as a sort of collective effect (valid in the low-energy regime) from this spacetime foam structure \cite{Rastgoo:2016syw}. The natural question thus arises as to how such a transition may take place and whether there might be any observable signatures of this foamy microstructure. The experimental signatures of \ac{QG} that are being searched at the moment and in the foreseeable future \cite{Bose:2017nin} are predictions of phenomenological toy models that try to capture only some aspects of quantum mechanics/quantum field theory and \ac{GR}, rather than robust predictions of a fundamental theory of \ac{QG} \cite{Carlip:2022pyh}.

\subsubsection{Some specific  approaches}

Condensed matter systems provide valuable lessons on how a geometric theory, featuring objects like curvature, torsion and nonmetricity, can arise as the continuum limit of some more fundamental theory, on which the metric and affine connection would represent some collective degrees of freedom. These geometric objects are crucial to capture the existence of microscopic defects \cite{Kittel2004}, which endow such systems with non-trivial topologies. Indeed, the discrete microstructure of certain materials (such as graphene) may behave in the continuum limit as emergent geometries, allowing the propagation of quantum fields on top of them  \cite{Iorio:2013ifa}. Gravitational theories with independent metric and affine structures naturally accommodate underlying foam-like structures and they can be excited via gravitational collapse \cite{Lobo:2014nwa}. At the effective level, one could build new gravitational theories based on the curvature/torsion/nonmetricity trio, such as $f(R)$~\cite{Sotiriou:2008rp}, $f(T)$~\cite{Cai:2015emx} and $f(Q)$~\cite{Harko:2018gxr}, connecting different implementations of this transition from the hypothetical foamy microworld to observational signatures in the macroworld.  

There are two other key influential concepts developed in the literature. The holographic conjecture based on quantum entanglement and emergent spacetime \cite{Susskind:1994vu,Maldacena:2013xja,Maldacena:2016hyu,Lulli:2021bme} and the entropic interpretation of emergent gravity~\cite{Verlinde:2010hp}. Other examples include continuous emergent microstructures  \cite{MacDowell:1977jt}, where gravity emerges from the symmetry breaking of a principal $\mathrm{SO}(4)$ bundle, or from a complex system of $N$ interacting particles with $\mathrm{O}(N)$ symmetry, which is governed by quantum mechanics 
\cite{Lee:2013dln}, or from gauge symmetry principles \cite{Wilczek:1998ea}. Spin foams \cite{Baez:1999sr} and fractal-based ideas \cite{Calcagni:2021ljs} also have their share in this discussion.

In minimal-length models, foamy effects come from the presence of a minimal accessible length \cite{Garay:1994en}, which modifies the Heisenberg or Poincar\'e algebra to accommodate a minimal uncertainty in position measurements at the Planck scale. In phenomenological models of quantum mechanics, the minimal length appears as a kinematic feature \cite{Bosso:2023sxr}, while the shape itself of the Hamiltonian may be deformed from the combined action of a modified position-momentum algebra and the choice of a relativity principle \cite{Bosso:2022rue,Arzano:2022har}. On the other hand, by equipping this framework to the standard quantum mechanical scenario with a stochastic nature, it is possible to mimic the foamy structure of spacetime in the non-relativistic regime and analyse potential experimental implications \cite{Petruzziello:2020wkd}.

\subsubsection{Some specific effects} 
\label{Sec:SpacetimefoamEffects}

Given the large variety of conceptual implementations of the transition of the spacetime foam to the macroscopic world, specific effects studied in the literature come in many shapes, of which we list several ones here:

\begin{itemize}
    \item Preferred frame effects, implementing deviations from the fact that the laws of Physics to be the same for all inertial frames, appear from theories implementing violations of Lorentz invariance, referred to as \ac{LIV} theories. They have been extensively tested \cite{Mattingly:2005re}.
    
    \item Time delays (\sref{ssec:TD}) and modified particle interactions (\ref{Sec:InteractionAnomalies}) may appear not only from \ac{LIV} theories, but also from \ac{DSR} theories, where Lorentz invariance is replaced by another symmetry principle (deformed Lorentz transformations) \cite{Amelino-Camelia:2002cqb} caused assuming a second observer-independent quantity (in addition to the usual light speed), namely, a high-energy scale, usually associated to the Planck scale.
    
    \item A non-trivial subluminal vacuum refractive index suppressed by the inverse Planck-scale mass \cite{Mavromatos:2009xg}. This can be caused by foamy effects that lead to a \ac{MDR} for particles and fields propagating on the foamy spacetime
        \begin{equation} \label{eq:dis}
            \bi{p}^2=F(E,m,\bi{p}^2,E_\text{P}) \,,
        \end{equation}
    where $E_\text{P}$ is the Planck length, $E$ the particle's energy, $m$ denotes a mass parameter, and the function $F$ encodes the \ac{MDR}. It be obtained from a generalized uncertainty principle \cite{Garay:1994en}, \ac{LIV} theories \cite{Mattingly:2005re}, \ac{DSR} theories \cite{Amelino-Camelia:2005zpp}, or more fundamental approaches with minimal length that lead to one of the phenomenological models~\cite{Freidel:2021wpl}. \Eref{eq:dis} represents a spontaneous breaking of the Lorentz symmetry by the ground state of foamy models.

    A competing theoretical mechanism that can cause a cosmic vacuum refractive index is associated with certain exotic models of dark matter~\cite{Gardner:2009et,Latimer:2013rja}. They can be distinguished from spacetime foam models through the specific dependence of the refractive index on the photon frequency and the emergence or non-emergence of birefringence~\cite{Ellis:2008gg}.
        
    \item Birefringence (different propagation speeds for each light polarization) from \ac{QG} spacetime with a  polymer-like structure at microscales in loop quantum gravity \cite{Gambini:1998it}. Such effect can be caused by mass-dimension five operators in the effective field theory extension of the QED sector of the Standard Model~\cite{Myers:2003fd}, by a non-minimal coupling between the electromagnetic potential and curvature from renormalization of QED on curved spacetime \cite{PhysRevD.22.343}, by the SME extension of the electromagnetic sector \cite{Kostelecky:2001mb} or by premetric electrodynamics in general \cite{PhysRevD.70.105022}. Birefringence effects are further discussed in \sref{Sec:PropagationEffects}.
    
    \item Stochastic effects in the velocity of light that lead to a spread of a photon pulse varying linearly with the photon frequency~\cite{Ellis:1999sd}, induced by space-time from $D$-particles. This effect needs to be contrasted to the corresponding effects induced by the \ac{QG}-induced refractive index \eqref{eq:dis}, which lead to a frequency-independent spread. Such effects can also be constrained by similar observations as the \ac{QG}-induced refractive-index effect.
    
    \item Alternatives to particle dark matter models in emergent gravity induced by modifications to the gravitational law via entanglement entropy  \cite{Verlinde:2016toy}.
    
    \item Induced triple \cite{Berglund:2022skk} and higher-order interference \cite{Berglund:2023vrm} in the context of ``gravitization of quantum theory'' \cite{Berglund:2022qcc}.
    
    \item Decoherence and breakdown of unitarity effects, particularly those leading to CPT violations \cite{Carrasco:2018sca} and neutrino decoherence from light-cone fluctuations \cite{Stuttard:2021uyw}.

    \item Broadening of spectral lines from foamy spacetime fluctuations \cite{Thompson:2006qe}.

    \item  \ac{QG}-induced stochastic \ac{LIV} effects of generic fuzzy space times~\cite{Vasileiou:2015wja}.
    
    \item Modifications of the waveform and the dispersion relation of gravitational waves due to polymer quantization, leading to a frequency dependence of the speed of the propagation of the waves~\cite{Garcia-Chung:2020zyq}.
    
    \item Variations of the speed of light, different dispersion relations between particles and antiparticles, and matter--antimatter lepton symmetry can be caused by $D$-brane models \cite{Alexandre:2007na} with bosons \cite{Ellis:1999uh} and fermions \cite{Ellis:1999sf} that lead to local effective spacetime metric distortions.

    \item Connection between \Ac{QG}  and the metaparticle concept \cite{Freidel:2018apz} as well as the concepts of Born geometry and Born reciprocity \cite{Freidel:2013zga}.

    \item Modifications to the properties of neutron stars such as masses-radius relations, limiting masses (e.g.\ white dwarf to neutron star) and moments of inertia (see the review \cite{Olmo:2019flu}), among others, features all accessible to different observational probes.

    \item Modified estimates on the dark matter budget due to the quantum fluctuations of the spacetime metric due to stringy foam-like structures involving spacetime $D$-particle defects \cite{Mavromatos:2010jt,Mavromatos:2010nk}.
\end{itemize}

On the other hand, specific effects of these models have been searched for using different observational channels:
\begin{itemize}  
    \item Using the Kilo-Degree Survey (KiDS) and the Galaxy And Mass Assembly (GAMA) survey \cite{Brouwer:2016dvq}, consistency of Verlinde's emergent gravity with weak gravitational lensing of low-redshift galaxies due to the displacement of dark energy by baryonic matter could be demonstrated.
    
    \item High-energy photons from, e.g., \ac{GRB} and blazar \cite{Li:2021gah} are used to constrain light speed variation with energy (suppressed by a power of the string mass scale) foam via \acp{MDR} of the form \eqref{eq:dis} \cite{Xu:2016zxi} and non-trivial refractive index \cite{Ellis:2008gg,Mavromatos:2009xg}, see also \sref{ssec:TD}. Constraints from birefringent effects and photon decays are consistent with this foamy scenario. Moreover, using specific \acp{GRB}, constraints on stochastic \ac{LIV} effects of generic fuzzy space times could be found exhibiting Planck-scale sensitivity~\cite{Vasileiou:2015wja}. These results set a benchmark constraint to be reckoned with by any \ac{QG} model that features spacetime quantization.

    \item Gravitational lensing and black hole shadows can be used to constrain a non-trivial refractive index of spacetime and \acp{MDR} \cite{Laanemets:2022rmn,Glicenstein:2019rzj,Barcaroli:2017gvg}.
    
    \item Null tests on foamy effects based on cumulative effects over distant sources from X-ray and \ac{GRB} observations \cite{Perlman:2014cwa}.
    
    \item Interferometric search for foamy effects, allowing to place constraints on specific implementations \cite{Ng:1999hm,Vermeulen:2020djm}. 

    \item Dedicated interferometers can in principle measure pacetime flusctuations that accumulate along the lightcone directions \cite{Verlinde:2019xfb}.

    \item The \ac{CMB} \cite{Planck:2015fie} and Big Bang nucleosynthesis  \cite{Planck:2018vyg} yield bounds on the baryon asymmetry that can be translated into bounds for $D$-brane models \cite{Ellis:1999uh,Alexandre:2007na,Ellis:1999sf}.
\end{itemize}

In summary, the spacetime foam idea, being a subset of the larger field of \ac{QG} phenomenology, suffers from the same fundamental difficulty, namely, the impossibility to directly test its effects at the Planck scale. There are, however, two generic effects that have been widely studied in the literature, namely, time delays in the propagations of cosmic messengers, and modifications of the kinematics of their interactions, which we discuss next in sections \ref{ssec:TD} and \ref{Sec:InteractionAnomalies} in more detail. We then identify possibilities how to detect these effects in future search strategies in \sref{sec:proposals}.

\subsection{Time delays in the propagation of cosmic messengers}\label{ssec:TD}

Time delays emerge already from the classical interaction between gravity and particles and fields propagating in vacuum that is described by the general relativistic geometry of spacetime. In this section, we briefly recall this classical achromatic vacuum time delay effects, that are already predicted by \ac{GR}, before examining additional energy-dependent (generally four-momentum dependent) time delays, due to microscopic quantum fluctuations or foaminess of spacetime (see \sref{S:stf}), which are widely discussed in the literature of \ac{QG} and its phenomenology. Then, we highlight the main challenges and difficulties in the search for such \ac{QG}-induced time delays.

\subsubsection{Classical time delays}
Even in the absence of \ac{QG} effects, the gravitational interaction causes vacuum time delays in the propagation of photons of any frequency and other cosmic messengers through spacetime, compared to their propagation in the absence of gravity. The most famous instance of this fact, which has been experimentally observed and actually constitutes one of the Solar-System tests of Einstein's \ac{GR}, is the Shapiro delay \cite{Shapiro:1964uw}: the observation that the travel time of a radar signal between two points in the vicinity of a gravitating object is increased compared to the travel time of the signal in the absence of the object. It is important to stress that the Shapiro delay is ``achromatic'', in the sense that it does not depend on the energy or polarization of the photon. In general, we gather large amounts of information about the gravitational interaction based on the observation of cosmic messengers propagating through spacetime and eventually being detect by us. Thus, in the pertinent data analyses of telescopes observing cosmic messengers, it is of utmost importance to take into account the classical -- and possibly the yet unknown additional \ac{QG}-induced -- vacuum time delay effects, in order to interpret the observations correctly. Famous examples where the classical time delays are already implemented in the data analysis are pulsar timing observations \cite{1976ApJ...205..580B,NANOGrav:2019jur} as well as cosmological 
redshift and distance observations \cite{Aghanim:2018eyx}. Pictorially, the time delay effects can be understood in the sense that gravity acts as a certain effective refractive medium for light propagation, thus introducing an effective refractive index \cite{perlick2004gravitational}. On the basis of Maxwell's electrodynamics and Einstein's \ac{GR}, the predictions for the classical time delays are independent of the energy (frequency) and of the polarisation of the photons, as a consequence of the Einstein equivalence principle \cite{Will:2014kxa}, which states that locally, that is, within an appropriately small region in the neighbourhood of each spacetime point, the weak equivalence principle and local Lorentz invariance holds, meaning that physics is described by \ac{SR}. In what follows we refer to this feature as \emph{local Lorentz symmetry}. Due to the universal coupling with gravity, such \ac{GR}-induced time delays do not only affect the propagation of light, but also all other standard model particles and fields including neutrinos~\cite{IceCube:2021tdn} and, due to its self interaction, gravitational waves~\cite{KAGRA:2013rdx,NANOGrav:2023gor}.

The above examples from classical gravity clearly demonstrate that (achromatic and polarization independent) time delays are an important prediction of \ac{GR}, which, on the one hand, allows us to test the theory, while on the other, gives us important information about the nature of the (classical) gravitational interaction, which is associated with the geometry of spacetime.

\subsubsection{Possible quantum gravity-induced time delays}
In addition to the general relativistic, achromatic and polarisation independent time delays discussed so far, time delays resulting from purely quantum effects of gravity might emerge. In contrast to the classical effects, they might depend on the energy and the polarisation of the probe, for the the following reason. Highly energetic particles and fields will interact with gravity on smaller length scales than low energetic ones. It is expected that the smaller the length scale which the particles and fields probe, i.e.\ the higher their energy, the more the particles and fields are affected by quantum vacuum fluctuations of gravity or spacetime foam, see also the discussion in \sref{S:stf}. Effectively, this scenario can be described as particles and fields propagating through a (refractive) medium generated by vacuum fluctuations of gravity, capturing aspects of a quantum spacetime or spacetime foam. The mathematical description of this pictorial idea has been further developed within various formal and phenomenological frameworks of \ac{QG}.

One intensively studied way to mathematically describe the interaction between particles and fields and the \ac{QG} ``environment'' (the effective medium or effective description of quantum spacetime) is to employ \acp{MDR}, which the four-momentum of the point-like particle excitations of the fields must satisfy.
Such \acp{MDR} can be determined from fundamental approaches to \ac{QG} by studying the propagation of fields (for example electromagnetic and gravitational waves), in string theory, loop quantum gravity, asymptotic safety or non-commutative geometry, it can be implemented phenomenologically \cite{Addazi:2021xuf} or be described geometrically \cite{Barcaroli:2015xda,Relancio:2020zok,Pfeifer:2021tas}.
 
A famous example of this type of \ac{QG}-induced time delays that is actively searched for, \cite{Terzic:2021rlx,Bolmont:2022yad}, is the delay in the arrival time (at a local observatory) of gamma rays of different energies~\cite{AmelinoCamelia:1997gz,Ellis:1999sd} which, however, are assumed to be emitted simultaneously, and at the same location in the universe. The predictions for the time delays differ depending on the particular \ac{QG} framework considered and \ac{MDR} employed (see \cite[Sec.~5.1]{Addazi:2021xuf} for an extended discussion). The first time delay derived in the context of \iac{MDR} of the form
\begin{equation}
\label{eq:MDRphotons}
E^2=\bi{p}^2\left[1\pm \left(\frac{E}{E_\text{QG}}\right)^n\right]
\end{equation}
for photons in a flat ($\Omega_k=0$) expanding universe with scale factor~$a(t)$, matter density~$\Omega_\text{m}$, and dark energy density~$\Omega_\Lambda$ \cite{Ellis:1999sd,Piran:2004ba}, predicts a time-of-arrival delay~$\Delta t$ between a low-energy and a high-energy photon emitted by a bursting source like a \iac{GRB} at redshift~$z$, as function of the energies~$E_0, E_1$ of the detected photons, to be
\begin{equation}\label{eq:timedelay}
    \Delta t_{\text{GRB-QG}} = \pm\frac{1+n}{2 H_0} \frac{(E_0^n - E_1^n)}{E_\text{QG}^n} \int_0^z \frac{(1+z')^n}{\sqrt{\Omega_\text{m} (1 + z')^3 + \Omega_\Lambda}}\,\rmd{z'}.
\end{equation}
Here $H_0$~is the Hubble parameter today, $E_{\textrm{QG}}$~the \ac{QG} energy scale, and $n$~the order of modification in the dispersion relation of photons. Since this first derivation, different time delay scenarios have been studied in the context for \ac{QG} phenomenology with Lorentz invariance deformation and violation \cite{Rosati:2015pga, Pfeifer:2018pty} and can be scrutinized by observations.

As in the classical coupling between gravity and the Standard Model fields, such energy-dependent time delays, possibly even of different magnitude, are also expected for messengers other than photons, such as neutrinos~\cite{IceCube:2021tdn, Carmona:2023mzs} and gravitational waves. An important point we would like to stress is that, despite the extreme weakness of \ac{QG} effects, nonetheless they could be significantly enhanced in cosmological situations, for example in searches for arrival time delays of probes that have travelled cosmological distances. Such huge distances can amplify  the extremely small quantum effects to a point of being potentially observable.

So far, \ac{QG}-induced time delays have not been observed in nature. Having said that, though, we also point out that there are no dedicated experimental setups optimized to search specifically for such a time delays.

\subsubsection{Theoretical challenges}
The aim to detect and interpret time delays correctly faces  theoretical and experimental challenges. Here we will discuss the theoretical ones, while the experimental difficulties will be discussed in section \ref{Sec:PropagationEffects}.

Energy-dependent time delays would imply the existence of deviations from local Lorenz invariance.
This would imply a modification of the causal structure of spacetime, together with a possible alteration of the notion of light-cones and the description of observers and their measurements. Each of these modifications or the conjunction of all of them may drastically impact the observability of any kind of time delay linked to the quantum regime of gravity. 

A particularly important challenge is the distinction between violation and deformation of local Lorentz invariance. In other words, either the \ac{LIV} implies the existence of a preferred reference frame, or the Lorentz group is replaced by the notion of spacetime and observer symmetry (\ac{DSR}). Both, the absence or the detection of any kind of time delay can be used as a guiding principle to identify the scale and the type of deviations from local Lorentz symmetry. Whether this happens effectively at an intermediate length/energy scale or at the fundamental \ac{QG} scale is unclear and depends on the theoretical approach under consideration. We would like to point out that for effects below the natural \ac{QG} scale it is also important to distinguish between \ac{QG} time delays and time delays due to a possible non-minimal coupling between curvature and electromagnetism, that might emerge due to renormalisation of QED on curved spacetime \cite{PhysRevD.22.343}.

It is often assumed that the \ac{QG} scale, at which time delay effects become important, is the Planck scale of~$E_\text{P} \approx 10^{28}~\mathrm{eV}$. However, such conclusions may be premature, since it is highly-model dependent. Even though some experiments exhibit sensitivities that for some processes exceed the Planck energy scale, the effective energy scale ($E_\text{QG}$) governing the time delays is often characterized by a combination of the Planck scale and further parameters of the theoretical \ac{QG} model under consideration. This has, for example, been pointed out in \cite{Ellis:2008gg,Ellis:2009vq} for certain models of spacetime foam stemming from string/brane theory (see \sref{S:stf}). There, the effective energy scale which characterizes such delays is a combination of the \ac{QG} (or string) mass scale divided by the linear (possibly redshift dependent) density of the spacetime defects encountered by the particle during its propagation. The observed time delays will also depend on the spacetime location of the source and one should consider combinations of searches at various redshifts, in order to constrain the effect properly and possibly separate it from other phenomena.

Yet further complications in the interpretation of possible time delays are systematic errors. Some of them might be related to the underlying cosmological model, which strongly influences the interpretation of the observed particle propagation. For instance, in modified gravity models that promote the cosmological constant~$\Lambda$ to a time-varying field to model dynamical dark energy, time delays are subject to systematic deviations from the classical $\Lambda$CDM~model. These may manifest themselves in the magnitude and type of time delays of messengers from sources and phenomena at high redshift such as \acp{GRB}, \acp{AGN}, or for strong lensing.

The interpretation of a detection or the absence of a time delay signal in gamma rays, gravitational lensing images, pulsar timing, neutrinos or gravitational waves from an astrophysical event of interest, requires the knowledge about competing effects which might mimic a propagation time delay, as discussed in section \ref{sec:expReq}. 

There are various competing models for the production of the different messengers. Additionally, classical propagation effects and time delays need to be taken into account. All of them depend on the underlying theory of gravity and the resulting cosmological model. These uncertainties about the source-intrinsic and classical propagation effects translate to less stringent constraints on the \ac{QG}-induced time delay of interest and pose a major challenge in the identification of a signal. However, a smoking gun signal, which is in principle capable of disentangling source-intrinsic from propagation time delays, is a redshift dependence of the expected delays predicted by different models.

If the \ac{QG}-induced time delays exist, they will not only manifest themselves in the observation of astrophysical sources but also in the interaction between gravity and quantum systems, such as Bose--Einstein condensates~\cite{Liberati:2005id,godtel2023constraints}. 
The advantage of using laboratory-sized quantum systems as sensors for \ac{QG} effects lies on the fact that they are highly controllable and can run over long time periods, which leads to high sensitivity. Such control is a qualitative improvement compared to the extraterrestrial cosmic observations with their inherited uncontrollable uncertainties regarding the state of the physical system under observation.

The challenge for detection based on quantum systems is to amplify the \ac{QG} effect over the time span of the experiment, such that it becomes detectable, or to increase the measurement sensitivity, in order to probe the desired effect.\medskip

In summary, it is known that gravity causes certain types of vacuum time delays. The search for additional, yet undetected energy-dependent time delays caused by potential \ac{QG} effects can be an important step in finding evidences of a theory of \ac{QG}. Thus, it is crucial to devise future dedicated experiments to address such a possibility both in an astrophysical context and in that of the more controllable Earth-based experiments. These two regimes complement each other, in the sense that they probe \ac{QG} time delays on different scales, which have different systematic uncertainties.

\subsection{Modifications of the kinematics of interactions of cosmic messengers}
\label{Sec:InteractionAnomalies}
Sufficiently energetic cosmic messengers, such as \acp{UHECR} and gamma rays, are expected to undergo interactions with the intergalactic medium affecting their observable properties. \Acp{MDR} of elementary particles could bring changes to the kinematics of these processes, for example by raising or lowering the minimum energy threshold above which an interaction is allowed by the conservation of four-momentum, introducing an upper threshold above which the process becomes forbidden, or even allowing new processes which in the Lorentz-invariant theory should be forbidden altogether \cite{Mattingly:2002ba}. The study of the appearance, disappearance, or shifting of the threshold energies of processes with respect to their values in \ac{SR} is called the study of ``anomalous thresholds''.

When studying threshold anomalies, it is very important to distinguish two scenarios: \ac{LIV} and \ac{DSR}. Both lead to different predictions.  While \ac{LIV} is usually only introduced at the level of in-vacuo dispersion relations, modified by a high-energy scale~$E_{\textrm{QG}}$ (usually taken as the Planck scale), \ac{DSR} scenarios supplement the \ac{MDR} by deformed relativistic symmetries, usually called deformed Lorentz transformations.  In the \ac{DSR} case, in order to ensure covariance of the energy and momentum conservation laws, these also undergo a compatible deformation. For example, the kinematics of a process~$A+B \to C+D$ are characterized by spelling out the \ac{MDR}s of \eref{eq:dis} and the energy and momentum conservation laws as
\begin{eqnarray}
    m_i^2 = E_i^2 - \bi{p}_i^2 + f(E_i, \bi{p}_i, m_i, E_{\textrm{QG}})\qquad(i=A,B,C,D),\\
    \bi{p}_A \oplus \bi{p}_B = \bi{p}_C \oplus \bi{p}_D, \qquad
    E_A \oplus E_B = E_C \oplus E_D,
\end{eqnarray}
where the function~$f$ specifies the in-vacuo \ac{MDR} and the $\oplus$~symbol describes the deformed four-momentum composition.  In the \ac{LIV} case, one would keep the four-momentum conservation as in \ac{SR}, i.e.\ the~$\oplus$ is simply the usual addition~$+$. All the deformations in the \ac{DSR} and \ac{LIV} cases disappear when the \ac{QG} scale~$E_{\textrm{QG}}$ goes to infinity (or, alternatively, in the limit where Lorentz invariance is fully preserved by \ac{QG}).

A most relevant example for phenomenological applications is electron--positron pair production from the interaction of very high-energy photons with low-energy photons, such as those from the \ac{CMB} and \ac{EBL}.
The \ac{LIV} and \ac{DSR} threshold energies $E_\text{th}$ for this process differ already at leading order, and can be written as deviation from the \ac{SR} threshold $E^\text{SR}_\text{th}$ as
\begin{eqnarray}
    E_\text{th}^\text{LIV} \approx \frac{m_e^2}{\epsilon} \left(1+\alpha\frac{m_e^4}{\epsilon^3E_{\textrm{QG}}}\right)
    = \left(1 + \alpha\frac{m_e^2}{\epsilon^2}\frac{E_\text{th}^\text{SR}}{E_{\textrm{QG}}}\right)E_\text{th}^\text{SR}, \\
    E_\text{th}^\text{DSR} \approx \frac{m_e^2}{\epsilon}\left(1+\beta\frac{m_e^2}{\epsilon E_{\textrm{QG}}}\right)
    = \left(1+\beta\frac{E_\text{th}^\text{SR}}{E_{\textrm{QG}}}\right)E_\text{th}^\text{SR},
\end{eqnarray} respectively~\cite{Carmona:2020whi},
where $\epsilon$~is the energy of the background photon and $\alpha$~and $\beta$~are real parameters characterizing the deformations (in the Lorentz-invariant case, $\alpha=\beta=0$). For positive~$\alpha$ and~$\beta$, the threshold is higher, so the universe is more transparent to high-energy radiation, while the converse holds when $\alpha$~or $\beta$ would be negative. The quantity $m_e^2/\epsilon^2$ (ranging from $\Or(10^{11})$ for visible \ac{EBL} photons to $\Or(10^{18})$ for \ac{CMB} photons) is an amplifier for the new physics effect, which makes the \ac{LIV} prediction much stronger than the \ac{DSR} one.  Therefore, while in the \ac{DSR} case the modification of the threshold is only appreciable when the energy is  close to the high-energy scale characterizing the correction to the kinematics, in the \ac{LIV} case the correction can be substantial at much lower energies. For this reason, only threshold anomalies in the \ac{LIV} framework are usually studied, except for the case of \ac{DSR} parameterized by a high-energy scale much lower than the Planck mass~\cite{Carmona:2021lxr}.

Threshold anomalies can be seen as kinematic modifications of the \ac{SR} description of interactions, and indeed, it is usually the only effect that is modelled to estimate the phenomenological consequences of \ac{LIV} or \ac{DSR} on the interactions of cosmic messengers~\cite{Martinez-Huerta:2020cut,Terzic:2021rlx}. However, the decay rates or cross sections of a process in \ac{LIV} or \ac{DSR} scenarios may differ from the standard ones even at energy ranges where the process is allowed in both, \ac{SR} and its modifications, which may be important for effects such as the possible modification of the transparency of the universe to high-energy gamma rays. The \ac{LIV} scenario has been explored in the past, but the availability of results for cross sections or decay rates is limited~\cite{Rubtsov:2012kb,Carmona:2022dtp,Carmona:2024thn}. For the \ac{DSR} scenario, changes in the cross sections or decay rates may only play a role in the case of a \ac{QG} energy scale much lower than the Planck one, comparable to the energies of cosmic messengers. While such dynamical calculations can be carried our for the \ac{LIV} case in the framework of effective field theory, the \ac{DSR} scenario still lacks a well-defined dynamical approach. In fact, an appropriate framework for \ac{DSR} should also address some of its known theoretical ambiguities, such as the soccer-ball problem~\cite{Hossenfelder:2007fy,Amelino-Camelia:2011dwc,Amelino-Camelia:2014gga, Amelino-Camelia:2020vvl, Kowalski-Glikman:2022xtr}, the spectator problem~\cite{Kowalski-Glikman:2004fsz,Carmona:2011wc,Amelino-Camelia:2011gae,Gubitosi:2019ymi}, or issues with momentum labelling~\cite{KowalskiGlikman:2002we,Kowalski-Glikman:2022xtr, Amelino-Camelia:2013sba}, whose resolution may require a novel perspective on \ac{DSR}~\cite{Carmona:2023luz}. Consequently, the implications of threshold anomalies in \ac{DSR} remain largely unexplored, for which it will be necessary to develop new approaches and techniques~\cite{Carmona:2021lxr}.
 
In the following we highlight specific hadronic, electromagnetic and weak interaction processes which are most promising to detect \ac{LIV} or \ac{DSR} threshold anomalies in the future. We briefly discuss the challenges that need to be attacked in order to detect these deviations from local Lorentz invariance.

\subsubsection{Hadronic processes (affecting cosmic rays)} 
\label{Sec:CosmicRaysInteractionAnomalies}

\Acp{UHECR} at the highest energies can undergo interactions with background photons, such as pion photoproduction and (for nuclei heavier than protons) photodisintegration.  
This introduces interaction horizons, i.e.\ a maximum distance from which cosmic rays above a given energy can be observed. Moreover, such interactions may produce fluxes of cosmogenic neutrinos and photons.  
Both these effects can be altered by \ac{LIV} in the hadronic sector \cite{Amelino-Camelia:2000bxx,Amelino-Camelia:2001com,Saveliev:2011vw,Martinez-Huerta:2017ulw,PierreAuger:2021tog}, so that \ac{UHECR} observations could in principle be used to test \ac{LIV} scenarios.
Unfortunately, as discussed in \sref{sec:AstrophysicalModelling}, several uncertainties about the sources and propagation of \acp{UHECR} complicate the interpretation of the \ac{UHECR} energy spectrum and mass composition at Earth in terms of those escaping their source environment.  Furthermore, the most recent such studies \cite{Aab:2016zth, AlvesBatista:2018zui, Heinze:2019jou, PierreAuger:2022atd} suggest that the spectral shape at the sources is harder than what is predicted by the Fermi mechanism, and that at the highest energies \acp{UHECR} are increasingly heavy nuclei, escaping their sources with Lorentz factors near or below the threshold for pion photoproduction. This substantially reduces their sensitivity to \ac{LIV} effects, as well as the production rates of cosmogenic neutrino and photons,  compared to the case of protons with the same energy.  
On the other hand, the presence of a non-negligible proton fraction at the highest energies can still not be fully ruled out by the most recent data, and it could be tested in the future with the aim of increasing the constraining power of the \ac{UHECR} spectrum and composition data, as will be mentioned in \sref{sec:proposals-UHECR}.

Another aspect to be explored regarding the hadronic sector is the development of the cascade of particles in the atmosphere after the first cosmic-ray interaction (extensive air shower), as discussed later in the final paragraphs of \sref{Sec:EffectsfromInteractionAnomalies}. 

\subsubsection{Electromagnetic processes (affecting gamma rays)} 
\label{Sec:GammaRaysInteractionAnomalies} 
A superluminal \ac{LIV} scenario would allow photon decay in vacuum; the detection of PeV-energy photons leads to very restrictive constraints on the \ac{LIV} energy scale for such a scenario \cite{LHAASO:2021opi}.  A subluminal scenario can lead to an increase of the energy threshold of the photon--photon interaction $\gamma \gamma \rightarrow e^- e^+$~\cite{Martinez-Huerta:2020cut}, and hence to an increase of the transparency of the universe to gamma rays, especially at energies $\gtrsim 10$~TeV, thus potentially allowing the detection of sources at greater distances (see e.g.\ \cite{Tavecchio2015rfa, Abdalla2018x}). However, using the current \acp{IACT}, no evidence for such \ac{LIV} effects has been detected. Such effects may be within reach of the future \ac{CTA}, with its significantly improved sensitivity and extended energy range, therefore providing unique tests of \ac{QG} phenomenology. 
That being said, one should keep in mind that \ac{CTA}, just as any other gamma-ray observatory, was not designed specifically to search for signatures of \ac{QG} as a primary goal, and that a dedicated experiment might be necessary to detect \ac{LIV}, or rule out a significant portion of the parameter space. 
At gamma-ray energies of a few hundred TeV and above, the target background radiation is the \ac{CMB}. In that regime, the anomalous absorption effect could be tested with water and hybrid Cherenkov detectors, such as \ac{HAWC}~\cite{historical:2023opo}, \ac{LHAASO}~\cite{Bai:2019khm}, or the planned \ac{SWGO}~\cite{Abreu:2019ahw}. The Pierre Auger Observatory is the most sensitive photon detector at EeV ($10^{18}$\,eV) energies \cite{PierreAuger:2022gkb}, and the non-observation of photons at such energies can be used to set limits on \ac{LIV} in the photon sector under certain assumptions about \ac{UHECR} sources \cite{Lang:2017wpe, PierreAuger:2021tog}.

\ac{DSR} scenarios do not allow photon decays, so limits on the decay rate cannot constrain them.
However, besides the time delays discussed in \sref{ssec:TD}, a more complex behaviour in the interaction of high-energy photons with the background can arise. For instance, one can find the expected flux to increase at lower energies but decrease at higher energies \cite{Carmona:2021lxr}.

Possible imprints of \ac{LIV} extensions of the Standard Model on the (inverse) Compton scattering effect is likely to be of paramount importance for the production of non-thermal gamma rays in the GeV--TeV range in many astrophysical sources, including \acp{GRB} and blazars.
Preliminary investigations show that such effects are plausibly only becoming discernible at $\sim$~PeV energies, where it could possibly lead to a suppression of Klein--Nishina effects \cite{Abdalla2018x}.
However, since at such energies the Compton cross section is already much smaller than the Thomson limit, this effect is unlikely to show any measurable signatures with astrophysical gamma-ray sources.

\subsubsection{Weak processes (affecting neutrinos)}
\label{anomalies-neutrinos}
In the standard \ac{SR} scenario, in subluminal \ac{LIV} scenarios and in \ac{DSR} scenarios, neutrinos can travel cosmologically long distances, being possibly subject to time delays, see \sref{ssec:TD} and \cite{Carmona:2023mzs}, with negligible probability of undergoing any interactions, but in a superluminal \ac{LIV} scenario sufficiently energetic neutrinos (e.g.\ with~$E \gtrsim 10~\mathrm{TeV}$ or~$100~\mathrm{PeV}$ for a linear or quadratic Planck-scale \ac{LIV}, respectively) can undergo weak decays in vacuum such as electron--positron or neutrino--antineutrino pair emission \cite{Carmona:2012tp,Carmona:2022dtp}. The rates of such processes can depend on the energy so strongly that the decay is nearly instantaneous for neutrinos above a certain energy and almost negligible below this energy. This effective threshold of dynamical origin may be the relevant one for the study of a particle reaction or a decay process, depending on whether it is higher or lower than the kinematic threshold.

For example, the kinematic energy threshold of electron--positron pair emission~($\nu\to \nu e^- e^+$) by superluminal neutrinos (with a MDR as Eq.~\eqref{eq:MDRphotons}, with the plus sign, in the massless limit), is~$E_\text{th} = (4 m_e^2 E_{\textrm{QG}}^n)^{1/(2+n)}$,
where $E_{\textrm{QG}}$~and $n$~are respectively the energy scale and the order that parameterize the \ac{LIV} correction, of order~$\mathcal{O}(E^n/E_{\textrm{QG}}^n)$ for a neutrino energy~$E$, whereas neutrino--antineutrino pair emission~($\nu\rightarrow\nu\nu\bar\nu$), also known as neutrino splitting, has a negligible threshold (zero in the limit of a massless neutrino).
The general expression of the decay width for these processes is~$\Gamma_{\nu_f \rightarrow X}(E) \propto E^5 (E/E_\text{QG})^{3n}$ \cite{Carmona:2022dtp}, where the proportionality coefficient depends on the initial neutrino flavour~$f$, the final-state particles~$X$, and on the order~$n$ of the correction.  This allows us to define an energy scale~$E_\text{th}^\text{eff}$ for each process such that~$\Gamma = H_0 (E / E_\text{th}^\text{eff})^{5+3n}$,
where $H_0$~is the present-day Hubble constant.
Owing to the large value of the exponent $(5+3n)$, the decay width is~$\Gamma\gg H_0$ for neutrino energies~$E > E_\text{th}^\text{eff}$ and~$\Gamma\ll H_0$ at lower energies, which is the reason why $E_\text{th}^\text{eff}$~acts as an effective dynamical threshold for the reaction to proceed. While this dynamical threshold is very relevant for the neutrino--antineutrino emission, in the case of the electron--positron emission the kinematic threshold is the most relevant one for values of the scale of the deformation below the Planck scale~\cite{ReyesHung:2022xut}.

In contrast, in a \ac{DSR} or subluminal \ac{LIV} scenario, no new effects in the propagation would become possible, but effects on the processes in which neutrinos are produced (such as pion decay) could still exist. For example, changes in the pion lifetime through \ac{LIV} effects could induce a modification of the muon neutrino spectra in IceCube \cite{Cowsik:2012qm}.

Note that neutrino splitting modifies the flavour composition of astrophysical neutrinos in the superluminal scenario, even if the \ac{LIV} term is diagonal in neutrino flavour space. On the other hand, non-diagonal \ac{LIV} terms affect neutrino oscillations in general~\cite{Arguelles:2021kjg}, modifying the flavour composition of astrophysical neutrinos at Earth. The search for effects in this direction has just begun recently~\cite{IceCube:2021tdn}. \medskip

In summary, models of \ac{QG} leading to \ac{LIV} and \ac{DSR} scenarios lead to modified threshold interactions, whose trace can be detected experimentally. It is important to study the different cases since they lead to different magnitudes of the effects. Often the \ac{DSR} effects are lower compared to the \ac{LIV} effects. A dedicated experiment searching for such effects should be able to distinguish the two cases.

\section{Astrophysical modelling and experimental requirements}\label{sec:expReq}

In this section, we discuss the experimental requirements for searches of signatures of \ac{QG} in astrophysical observations. A definitive feature of research using astrophysical observations is the impossibility of controlling the source of the signal and the medium between the source and the detector. The limited knowledge of these is an important source of systematic uncertainties. 
Moreover, there are other competing effects which could mimic or deprecate new physics effects. Some of these are known , e.g.\ secondary particles which are not emitted directly from the source yet cannot be resolved from the primary particles. Others could be hypothetical different parts of new physics affecting each other, e.g.\ new particles effects such as interactions, productions or decays involving axion-like particles. 
The importance of modelling of astrophysical sources, and the medium through which messengers propagate is discussed in \sref{sec:AstrophysicalModelling}. Specifically we focus on the modelling of their emission in \sref{sssec:modelling}, their interaction with background radiation fields like \ac{EBL} and \ac{CMB} in \sref{sssec:backgroundrad}, their interaction with astrophysical obstacles in \sref{sssec:interactionback} and their interaction with possible further intergalactic impediments in \sref{sssec:convent}. We also discuss the uncertainties introduced by these, as well as how those uncertainties might be reduced.

Another essential ingredient is the advancement of experimental setups. This could be achieved through improvements of existing observational equipment or techniques, or by introducing entirely new measurement and analysis methods. Experimental requirements for performing and improving the searches for signatures of \ac{QG} are discussed in \sref{Sec:SpecificRequirements}. In \sref{Sec:PropagationEffects} we focus on the detection of effects the messengers experience while propagating through vacuum. Namely, we discuss time delays, as introduced in \sref{ssec:TD} and birefringence caused by \ac{LIV}. In \sref{Sec:EffectsfromInteractionAnomalies} we discuss requirements for detecting effects resulting from interaction anomalies. 
The detection of \ac{QG} effects in diffuse fluxes of UHECRs, photons, and neutrinos, as well as in neutrino oscillations, and during the development of atmospheric showers, each require different experimental specifications for their detection and are, therefore, discussed separately.

This white paper, while aimed at a wider audience, was also composed by researchers working in different fields of research of \ac{QG} phenomenology. Through the process, we realised that certain terms, commonly in use in one research field, are not as clear, or even have a different meaning in other fields. Therefore, we explain the meaning of the following two notions:

\textbf{Sensitivity to some observable} indicates how accurately and precisely some observable can be measured with a certain instrument or method. In the case of \ac{QG} energy scale, which has not been detected and only lower limits are being set, the sensitivity means how much of the parameter space certain method can cover.

\textbf{Instrumental sensitivity or detector sensitivity.} In astrophysics, it refers to how low the source flux can be in order to be detected by a given detector within certain observation time. A more sensitive instrument will take less time to detect some source. Putting it the other way around, a more sensitive instrument will collect more signal within the same observation time.

\subsection{\label{sec:AstrophysicalModelling}Modelling of astrophysical sources and the propagation of messengers}

To investigate the potential effects of \ac{QG} using cosmic messengers, it is crucial to understand how they are produced and how they propagate. 
Improving our ability to model the processes within the sources of cosmic messengers, as well as those affecting their propagation, is paramount for increasing the sensitivity of \ac{QG} experiments. 
Ideally, we would need to know the precise energy and emission time of every single particle at the source. This is, of course, unfeasible. However, a precise multiwavelength modelling of the source-intrinsic flux, including its polarisation properties and its flavour content, as well as spectral correlation between different energy bands, would significantly increase the sensitivity to the \ac{QG} energy scale. 

A notable case are \acp{GRB}, whose long-duration population generally feature spectral lags, in contrast to the sort-duration ones~(see e.g.\ \cite{Dai:2017dzh} and references therein).
This example clearly demonstrates the importance of a close collaboration between astrophysicists working on source modelling and the broad \ac{QG} phenomenology community.

\subsubsection{Uncertainties from the modelling of source emission processes}\label{sssec:modelling}
In \cite{Perennes:2019sjx}, a detailed study of intrinsic spectral lags in flaring \acp{AGN} above 100~GeV was conducted in order to investigate whether these can be distinguished from \ac{LIV}-induced time delays. The authors concluded that, while certain intrinsic spectral lags are to be expected, their magnitude can vary in time. This means that, if the intrinsic time delay between two photons of energies $E_h$ and $E_l$ is modelled as
\begin{equation}
    \Delta t \propto (E_h^\alpha - E_l^\alpha),
\end{equation}
the exponent~$\alpha$ will not be constant in time. This is contrary to the expected \ac{QG}-induced effects, as displayed in Eq.~\eref{eq:timedelay} in \sref{ssec:TD}.
Although certain classes of objects like GRBs and blazars do behave differently across cosmic history \cite{Hasinger:2005sb, Hopkins:2006bw, Wanderman:2009es}, for most of the analysis it is assumed that source-intrinsic time delays do not depend on the source distance from Earth.

Precise spectral modelling at the highest energies is also crucial for testing anomalous absorption by the background fields, as described in sections~\ref{Sec:CosmicRaysInteractionAnomalies} and~\ref{Sec:GammaRaysInteractionAnomalies}. In~\cite{Lang:2019PhRvD}, a \ac{LIV} analysis was made on 18 spectral measurements from 6 different \acp{AGN}. In all cases a power law with exponential cutoff was assumed as the shape of intrinsic spectra. The authors investigated the systematic uncertainty introduced with this choice by also assuming a power law spectral distribution and comparing the results. 
While an exponential cutoff did lead to better fits to the data, it also yielded more conservative results,
reducing by a factor $\lesssim 2$ the lower limit on the \ac{QG} energy scale, compared to the power law result,
which is similar to the difference between using different exponential cutoff models. In~\cite{Abdalla_HESS_Mrk501:2019ApJ} only one sample from the \ac{AGN} Mrk\,501 was analysed. In this case, a power law was assumed as the best choice for the intrinsic spectral shape, even though this sample was one of the spectra used in~\cite{Lang:2019PhRvD}. Furthermore, a source-intrinsic spectral upturn could mimic \ac{LIV}-induced modified gamma-ray absorption. According to \cite{Abdalla_HESS_Mrk501:2019ApJ}, this is the main source of uncertainty on the \ac{QG} energy scale. Knowing the correct spectral shapes would remove these ambiguities and degeneracies.

\subsubsection{Uncertainties from background radiation fields}\label{sssec:backgroundrad}
Understanding the propagation of messengers entails considering various matter and radiation fields that could act as targets, including the \ac{CMB}, \ac{EBL}, and interstellar radiation field. Uncertainties in the \ac{EBL} spectrum can significantly influence the propagation of gamma rays~\cite{Franceschini:2008tp, Dominguez:2010bv, Saldana-Lopez:2020qzx} and \acp{UHECR}~\cite{Batista:2015mea, AlvesBatista:2019rhs}, thereby impacting the observables relevant to any potential \ac{QG} signals. Studies taking into account the \ac{UHECR} propagation through the extragalactic space for reproducing the measured energy spectrum and mass composition at Earth show differences up to $3.5\sigma$ among scenarios involving different \ac{EBL} models, if the standard propagation is considered~\cite{Aab:2016zth}. The spatial distribution of the \ac{EBL}, despite being fairly homogeneous and leading to discrepancies $\lesssim 1\%$ in the propagation of gamma rays~\cite{Abdalla:2017chl}, also represents another source of uncertainty. 
Fortunately, with advancements in observational techniques across different wavelengths and the employment of multiple strategies, it is anticipated that the uncertainties associated with the \ac{EBL} will be mitigated in the near- and mid-future, even for high redshifts ($z \sim 6$)~\cite{Saldana-Lopez:2020qzx}. Upcoming \acp{IACT} such as \ac{CTA}~\cite{CTAConsortium:2018tzg} will contribute to improving this picture by a factor between 2 and 3, with respect to current facilities of this kind, for at least 800~hours of observations of selected \acp{AGN}~\cite{CTA:2020hii}.

Note, however, that Galactic sources such as pulsars are not affected by uncertainties in the \ac{EBL}. In this case, the main uncertainties relate to their distance, which may vary by up to 30\%~\cite{Ahnen:2017wec}. This could possibly be reduced to $\sim 15\%$ thanks to recent parallax measurements from GAIA~\cite{2023A&A...674A...1G}. 

Another source of scattering targets comes from infrared radiation by dust~\cite{Vernetto:2016alq}, adding the interstellar radiation field to targets from \ac{EBL}. However, this is a secondary effect for extragalactic sources, since the distance that the signal travels through the Milky Way is orders of magnitudes shorter compared to the overall distance from a source. In fact, even the scattering of signal from Galactic sources is considered negligible in present-day experiments. Given a relatively short distance and consequential low scattering probability, the effect is subdominant compared to the systematic effects of current facilities. Nevertheless, the interstellar radiation field may play an important role for Galactic sources, especially for sources emitting gamma rays with energies $\gtrsim 100 \; \text{TeV}$~\cite{Porter:2017vaa}. 

\subsubsection{Uncertainties from interaction cross sections with background and magnetic fields}\label{sssec:interactionback}
The knowledge of cross sections for the corresponding interactions with various astrophysical targets
is relevant in the search for new physical effects. 
This is particularly problematic for \acp{UHECR}, since photonuclear cross sections are poorly known at these high energies~\cite{Khan:2004nd, Batista:2015mea, Boncioli:2016lkt, Soriano:2018lly} and can lead to discrepant interpretations of the observations~\cite{Allard:2005ha, Aab:2016zth, AlvesBatista:2018zui, Heinze:2019jou}. In~\cite{Aab:2016zth} the changes in the spectral parameters for \acp{UHECR} at the escape from the sources due to the different cross section models as well as to the different \ac{EBL} models can produce differences at the level of $2\sigma$ in the fit of the \ac{UHECR} spectrum and composition at Earth. These cross sections and their thresholds are the ones that change in the presence of \ac{LIV}, for instance, both for \acp{UHECR}~\cite{Saveliev:2011vw, Lang:2017wpe, Anchordoqui:2017pmf} and gamma rays~\cite{Martinez-Huerta:2020cut}. In particular, since \ac{LIV} modifications increase the threshold for photonuclear cross sections, photodisintegration processes are less impactful for large \ac{LIV} parameters~\cite{PierreAuger:2021tog}. Therefore, in order to be sensitive to small deviations from SR, accurate knowledge of photonuclear cross sections and of the \ac{EBL} are crucial to properly identify \ac{QG} signatures using high-energy messengers.

Magnetic fields within the Milky Way and beyond can alter the trajectories of charged particles. While \acp{UHECR} are particularly susceptible to these effects~\cite{AlvesBatista:2017vob, Hackstein:2017pex}, neutrinos are not, making them reliable messengers for studies requiring directional accuracy. Gamma rays are not directly affected by magnetic fields, but the charged component of the electromagnetic cascades that they induce might be~\cite{Aharonian:1993vz, Plaga:1995ins, Neronov:2009gh, Batista:2021rgm}. Although some constraints exist regarding the distribution of cosmic magnetic fields in regions such as galaxies and galaxy clusters, their characteristics in the filaments connecting clusters are less well-known~\cite{Ryu:2011hu, Vazza:2017qge}. In cosmic voids, which dominate most of the volume of the universe, knowledge about magnetic fields is even more limited~\cite{Vachaspati:2020blt}. This directly translates into uncertainties in spectral, temporal, and directional observables. Future polarization surveys such as the Square Kilometre Array \cite{SKAMagnetismScienceWorkingGroup:2020xim} and the Next Generation Very Large Array \cite{Lacy:2019rfe} are expected to reduce these uncertainties by increasing sample size by roughly one order of magnitude. The Galactic magnetic field, in particular, will likely be much better modelled using the upcoming data and new computational tools such as IMAGINE \cite{Boulanger:2018zrk}.
Fields in cosmic voids, however, will hardly be measured in this way, but gamma-ray measurements are expected to deliver better constraints~\cite{CTA:2020hii}. 

If the fields are weak ($\lesssim 10^{-17}~\mathrm{G}$), their impact on time delays and arrival directions of astroparticles should be small; if they are strong ($\gtrsim 10^{-14}~\mathrm{G}$), this might compromise time-delay studies. For \acp{UHECR}, the average delay is typically $\sim 1.5 Z~\mathrm{yr}$ for a nucleus of atomic number $Z$ with $100~\mathrm{EeV}$ coming from a source $10~\mathrm{Mpc}$ away, assuming a magnetic field of strength $B \sim 1~\mathrm{pG}$ and coherence length of $1~\mathrm{Mpc}$. This time delay scales roughly with the inverse of the energy squared, such that  a $10~\mathrm{EeV}$ proton from would be delayed by over a millennium for the same magnetic field~\cite{Waxman:1996zn}. Therefore, \acp{UHECR} are generally not reliable messengers for time delay studies. Gamma rays, on the other hand, can be more reliable probes, since they could have originated from an electromagnetic cascade whose short-lived charged component is affected by magnetic fields only over scales comparable to those of inverse Compton scattering ($\sim 100~\mathrm{kpc}$). In this case, gamma rays observed with energies of 100~GeV from nearby sources ($z < 0.05)$ would have time delays of $\lesssim 1~\mathrm{day}$ for $B=10^{-18}~\mathrm{G}$, whereas for a source at $z \simeq 0.5$ this would increase by a factor 5~\cite{Neronov:2009gh}. Because the time delays scale with the square of the magnetic field, for $B = 10^{-15}~\mathrm{G}$, they would exceed a century even if the sources are closeby~\cite{Batista:2021rgm}. In the case of gamma-ray observations, these uncertainties could indeed dilute the flux from a short-duration \ac{GRB} over much larger time scales~\cite{Takahashi:2010cg} and affect inferred intrinsic source properties~\cite{Saveliev:2020ynu}. 

\subsubsection{Further uncertainties from propagation through intergalactic space}\label{sssec:convent}
Other conventional phenomena might also be at play and affect the propagation of cosmic messengers. A potentially important one that might affect gamma-ray--induced electromagnetic cascades in intergalactic space are plasma instabilities, stemming from the interaction of the electrons with the medium. The role of this effect is far from clear~\cite{Broderick:2011av, AlvesBatista:2019ipr, Perry:2021rgv} and can hardly be assessed without detailed particle-in-cell simulations, which is not feasible for low-density environments such as the intergalactic medium. This could compromise the interpretation of gamma-ray observations and change the ratio between the fluxes of \acp{UHECR} and the corresponding cosmogenic photons. 

\Acp{UHECR} produce neutrinos and photons of cosmogenic origin during intergalactic propagation~\cite{Kotera:2010yn, AlvesBatista:2018zui, Heinze:2019jou}, which can act as backgrounds when studying individual astrophysical objects. Similarly, uncertainties inherent to the production of neutrinos and photons through cosmic-ray interactions in the large-scale structure of the universe such as galaxy clusters can act likewise~\cite{Berezinsky:1996wx, Fang:2017zjf, 2021MNRAS.507.1762H, Hussain:2022tls}. Evidently, this can be mitigated by reducing uncertainties related to their progenitor cosmic rays, which is expected to happen with future \ac{UHECR} facilities~\cite{Coleman:2022abf}. 

All the modelling uncertainties listed above have to be accounted for in searches for \ac{QG} phenomena. This is, however, computationally challenging, and requires advanced software tools to scan the complete parameter space efficiently, including both conventional phenomena and \ac{QG}-related ones. Existing codes, such as CRPropa~\cite{Batista:2016yrx, AlvesBatista:2022vem} (cosmic rays, gamma rays, neutrinos), \textit{SimProp}~\cite{Aloisio:2017iyh} (cosmic rays and neutrinos), \texttt{ELMAG}~\cite{Blytt:2019xad} (gamma rays), GALPROP~\cite{Porter:2021tlr} (Galactic cosmic rays), among others, can be adapted for this purpose, as done in~\cite{Reyes:2023osq, Saveliev:2023efw}, ultimately leading to better models for interpreting observations. Nevertheless, it remains unclear how to \emph{efficiently} sample from this vast multi-dimensional parameter space. Some proposed strategies include, for example, using machine learning methods or deep probabilistic programming~\cite{AlvesBatista:2021eeu}.

\subsection{Experimental requirements}
\label{Sec:SpecificRequirements}

In what follows, we discuss the experimental requirements needed to improve the sensitivity of specific telescopes which are used to find some of the effects discussed in \sref{sec:WhatEffects}.  \Sref{Sec:PropagationEffects} deals with the requirements to measure propagation effects: time delays and birefringence (described in sections~\ref{ssec:TD} and~\ref{Sec:SpacetimefoamEffects}, respectively), while \sref{Sec:EffectsfromInteractionAnomalies} considers \ac{LIV} and \ac{DSR} effects coming from the interaction anomalies discussed in \sref{Sec:InteractionAnomalies}, which modify the expected fluxes of \acp{UHECR}, gamma rays, and neutrinos. We will also considers effects in the development of showers in the atmosphere, which are relevant in the detection of some of the cosmic messengers.

\subsubsection{Propagation effects in vacuum}
\label{Sec:PropagationEffects}
In this section, we will consider the two main consequences that can be associated to propagation effects: the existence of chromatic \ac{QG}-induced time delays (\sref{ssec:TD}), which may be observable in the study of the arrival times of photons and/or neutrinos, and birefringence, an effect for photons predicted in specific frameworks within effective field theory, as indicated in \sref{Sec:SpacetimefoamEffects}. Notice that  time-of-flight studies concerning gravitational waves are not sensitive to Planck-scale physics, given the very low energies involved \cite{Cornish:2017jml}.

\paragraph{Photon time delays.}
Time delays are one of the most prominent \ac{QG} phenomenology effects, as discussed in \sref{ssec:TD}.
Testing the energy dependence of the speed of photons relies on comparison of the detection time of individual highly energetic photons with that of lower-energy photons. The influence of \ac{QG} effects on low-energy photons is usually neglected, meaning that their temporal distribution at the detection is assumed to be the same as at the emission; that of high-energy photons, on the other hand, is modified by the effects of \ac{QG}.
The systematic uncertainty in the detection times of individual photons (regardless of their energy) is usually negligible, as they are measured with \textmu{s}~accuracy. 

The temporal distribution of emitted particles is usually described with a light curve template defined by several parameters. 
The magnitude of uncertainties of these parameters varies from case to case, and is a matter of compromise between several factors. Faster and more pronounced flux variability puts stronger constraints on the emission time: the sensitivity to the \ac{QG} energy scale is inversely proportional to the variability time scale for linear ($n=1$) \ac{QG} effects and to its square root for quadratic ($n=2$) ones \cite[Table~1]{Terzic:2021rlx}. 
This, however, requires using shorter time bins, which results in relatively smaller number of events per bin, and, consequently, larger uncertainties. For this reason, brighter, more variable emissions are selected to study time delays. Improvement of instrumental sensitivity will directly reflect on the ability to model the emission light curves, and on the sensitivity to the \ac{QG} energy scale. 

The effect of time delay is linearly dependent on the photon energy both for $n=1$ and $n=2$~\cite{Terzic:2021rlx}. However, it is not enough to have high-energy photons in the sample. Since lower-energy photons are used as references, with the effects of \ac{QG} usually neglected, and \ac{QG} is expected to affect higher-energy photons more, a favourable sample will include photos with a wide range of energies. Having a narrow energy range sample, and assuming that low-energy photons are not affected by \ac{QG} will introduce a strong bias towards the absence of time delay. In addition to wide energy rage, which is a characteristic of a sample, one should aim at good energy resolution, which is a characteristic of a detector, because large uncertainties in the energy reconstruction are propagated to the analysis, again reducing the ability of detecting any time delay.

Another critical factor for sensitivity to the \ac{QG} energy scale, related to instrumental sensitivity, is the source distance. In~\cite{Terzic:2021rlx}, the sensitivity to the \ac{QG} energy scale depends linearly on the redshift of the source (for extragalactic sources) for $n=1$, and is proportional to the redshift to the power of $2/3$ for $n=2$. 
However, gamma rays with energy higher than $\sim 100$~GeV are absorbed on background electromagnetic radiation, decreasing the total number of gamma rays reaching the detectors. Improving instrumental sensitivity cannot affect absorption on background radiation, but it can increase the rate of gamma rays detected among the ones that survive. However, it is important to mention that closer-by sources such as low-redshift \acp{AGN} and Galactic pulsars provide very competitive sensitivities and limits, as they are not affected by gamma-ray absorption at energies $\lesssim 100 \; \text{TeV}$.

According to~\cite{Bolmont:2022yad}, systematic effects related to the light curve uncertainty spoil the sensitivity to the \ac{QG} energy scale by a factor of $\sim 2.3$ for linear or $\sim 1.5$ for quadratic correction of the dispersion relation. In most cases for $n=1$, the dominant systematic effect comes from the uncertainty of the light curve template parameters. This is true for most of the individual sources considered, as well as for certain type of sources combined or combination of all sources used in the study. Indeed, the variance of the template statistics constitutes more than 50\% of the total systematic variance. For $n=2$ this falls down to about 20\%, and the contribution of the template statistics to the total systematic variance is somewhat less, but still comparable to some other sources of systematic uncertainty, such as energy scale uncertainty. 
In the case of pulsars, that uncertainty is greatly reduced, since only one pulse shape needs to be understood, instead of a series of complicated pulses, which reduces that systematic uncertainty to considerably less than 10\% \cite{Ahnen:2017wec}. Nevertheless, modifications of pulse behaviour in the transition from curvature radiation to inverse Compton scattering of photons between 200 and $300~\text{GeV}$ have been observed~\cite{Ahnen:2017wec}. Several competing models for very-high energy pulsar emission have been proposed~\cite{Aharonian:2012zz, Hirotani:2014gxa}, which do include intrinsic spectral lags. Observational discrimination between these pulsar models is hence desirable for \ac{LIV} searches using this technique.

Additionally, the assumption that the emission times in different energy ranges have the same distribution is a rather strong one, especially considering that intrinsic energy-dependent time delays have already been unambiguously detected in some \acp{GRB}~\cite{Dai:2017dzh}. 

Besides a precise broadband modelling of the source-intrinsic flux discussed in \sref{sec:AstrophysicalModelling}, one could try developing an analysis method which does not rely on making assumptions on the distribution of events at emission.

As explained above, a specific implementation of the likelihood method that has been used so far requires a parameterization of the low energy light curve, which is the biggest source of systematic errors. 

An improvement would be to find a way to implement a likelihood method directly from energies and times of detected photons, without the need to provide parametrized temporal and spectral distributions.

Several analysis methods, such as the energy cost function~\cite{Albert:2007qk}, the sharpness maximization method~\cite{Vasileiou:2013vra, deAlmeida:2012cx}, the dispersion cancellation~\cite{2008ApJ...673..972S}, the minimal dispersion~\cite{Ellis:2008fc}, among others, are based on removing the dispersion from the sample by maximizing the sharpness of the light curve. They do not depend on the explicit temporal distribution of events at the emission, and are, therefore, free of making assumptions. However, treatment of systematic uncertainties, in particular the ones coming from instrumental effects, is not as transparent as in the likelihood approach. Moreover, these methods do not allow for a simple consideration of multiple sources in a single analysis. Retaining the likelihood approach, but removing the necessity of assuming a particular distribution of events at emission seems to be the optimal path. A so-called binned likelihood method has been presented in \cite{Abe:2023by}. There is no assumption on the temporal distribution of events at emission, consequently the data sample is not split in the low and high energy bands, and the approximation of negligible influence of \ac{LIV} on low energy events is not necessary. However, it is still not clear whether the method implicitly assumes equal distributions of events in lower and higher energies. Although a proper scrutiny of the method is yet to be made, it is expected that the binning of the data will make the method somewhat less sensitive to \ac{QG} effects, but more robust. 
To summarise, improvements of the sensitivity of time of flight studies on gamma rays can be achieved by:
\begin{itemize}
    \item enhancing detector sensitivity, which increases the statistics of samples, which in turn enables more precise emission modelling;
    \item improving energy resolution and widening the energy range;
    \item introducing novel analysis methods that are less dependant on assumptions about the distribution of events at emission.
\end{itemize}

\paragraph{Neutrino time delays}
While having at the moment a poorer statistics with respect to photons, neutrinos are potentially the most sensitive messengers for time delay studies. Unlike photons, neutrinos can travel cosmologically large distances with negligible absorption at all energies, assuming no relevant changes in threshold reactions due to \ac{LIV} or \ac{DSR}, which are discussed next in \sref{Sec:EffectsfromInteractionAnomalies}. Moreover, source intrinsic effects are not expected to be as important as in photon time delay studies, given the considerably higher energies [$\mathcal{O}(\text{PeV})$] reached by astrophysical neutrinos, which give much longer time delays than in the photon case.

For neutrinos of energy $\sim 100~\mathrm{TeV}$ coming from a source at redshift $z\sim 1$, the expected time delay with respect to low-energy signals is, according to~\eref{eq:timedelay} and assuming massless neutrinos, of the order of one day. For neutrinos in the PeV energy range one might expect delays of the order of tens of days.

Currently, the largest and most sensitive neutrino telescope is IceCube, which has observed  neutrinos of extragalactic origin in the energy range from $60$~TeV to a few PeV \cite{IceCube:2020wum, Aartsen:2013bka, Abbasi:2020jmh,Aartsen:2016xlq,Stettner:2019tok}.
This lower bound on the energy  comes from the need to disentangle the neutrino signal of astrophysical origin from background signal, such as atmospheric neutrinos \cite{IceCube:2020wum}.
The rate of these events is around 15 candidate astrophysical neutrinos per year \cite{IceCube:2020wum}, which is too low to detect more than one high-energy neutrino from a single emission event. 
Therefore, time delay studies cannot rely on the comparison of the time of flight of neutrinos with different energies. Rather, they rely on the comparison of the time of flight of one neutrino with a low-energy electromagnetic counterpart. So far, studies have focussed on the association of neutrino signals with \acp{GRB}, given that the latter have a  time span smaller than the expected time delay effect. The identification of the electromagnetic counterpart of the neutrino signal is done  by means of a directional correlation criterion. Note that one cannot associate unambiguously a given source to individual neutrino events, due to the directional uncertainty and the large neutrino energy, resulting in a large time delay expected between the neutrino and the lower energy photons. In fact,  excluding the presence of any time delay effect, at the moment no associations between a high-energy neutrino and a \ac{GRB} have been found.  Therefore, for time delay searches, the directional association is done statistically over the whole population, similarly to what is done for searches that do not account for time delays \cite{IceCube:2022rlk}. 
Once the possible set of associations is identified, time delay effects are searched for by looking at the correlation between the neutrino-\ac{GRB} time delay, the neutrino  energy and the redshift of the source, according to \eref{eq:timedelay}.

Directional association is affected by the uncertainties on the neutrino reconstructed direction. Larger uncertainties result in a larger background, namely a larger number of accidental associations. Assuming a uniform distribution of \ac{GRB}s in the sky, the expected number of background events scales linearly with the area covered by the direction uncertainty.
Identification of the electromagnetic counterpart of the neutrino signal is also needed in order to pinpoint the redshift of the source, and thus estimate the distance travelled. The redshift measurement is currently the main source of uncertainty in these time-of-flight studies, due to the fact that for most of the \acp{GRB} the redshift is unknown \cite{grbcatalogue}. 

Over a large population, redshift assignment can be done statistically, based on the sources for which the distance is known \cite{Amelino-Camelia:2022pja}. However, this might introduce a bias, given that the redshift distribution of \acp{GRB} with known redshift might be different from the one of \acp{GRB} with unknown redshift. For this reason, neutrino time of flight analyses would benefit greatly from improved capabilities of measuring the redshift of \acp{GRB}.

The reconstruction of the neutrino energy is based on the deposited energy in the detector. Neutrino events are usually classified as track or cascade events depending on the topology of the interaction vertex inside the detector. Charged-current muon-neutrino events have a muon as the leading visible outgoing particle in the first interaction. Since muons traverse a long distance in the ice, these are called \emph{track events}. 

These kinds of events have a good angular resolution (about $1^\circ$  above TeV energies \cite{IceCube:2020wum}), but the neutrino energy reconstruction is affected by large uncertainties, of about $30\%$ \cite{IceCube:2020wum}, given that the event can start outside of the detector.
For \emph{cascade events} the interaction does not produce visible muon tracks. Their energy is contained completely in the ``instrumented volume'', so that the energy reconstruction of the neutrino is very accurate, giving an energy uncertainty of about $10\%$ \cite{IceCube:2020wum}. However, the topology of the interaction makes it harder to determine the incoming direction of the neutrino, so the angular resolution of these events is much lower with respect to tracks (of the order of $10^\circ$). It is then clear that there is a trade off between energy and angular resolution: shower events optimize energy resolution at the expense of angular resolution, while the converse holds for track events. While current studies choose to focus on just one kind of events in order to have a homogeneous sample, accounting for the full probability distribution function of energy and angular information will enable all-event studies in the future. 

Finally, a consistent LIV analysis of neutrino time delays should also consider the interplay between time-of-flight anomalies and the instability affecting superluminal neutrinos, as explained in section~\ref{anomalies-neutrinos}, which imposes strong constraints on the existence of early neutrinos exceeding an energy determined by the LIV scale~\cite{Carmona:2023mzs}.

The main requirements to improve the sensitivity of time of flight studies that use astrophysical neutrinos are the following:
\begin{itemize}
\item precise estimate of the redshift of the sources;
\item improved energy reconstruction techniques for track events;
\item improved techniques for reconstructing the direction of cascade events;
\item increase the rate of observed neutrino events, so to allow for neutrino-neutrino time of flight studies;
\item consistently introduce neutrino anomalous decays in the analyses of time delays.
\end{itemize}

\paragraph{Birefringence} 
\Ac{LIV} effects could introduce modifications to the phase and group velocities of the circularly polarized modes of photons. When these modes have different velocities, propagation of light is subject to vacuum birefringence, which induces a rotation of the polarization direction of linearly polarized signals, or a loss of polarization degree~\cite{Contaldi:2008yz, Maccione:2008tq, Toma:2012xa, Wei:2019}.
To observe this effect, knowledge of the emission properties of the photon source is required. Moreover, since the degree of polarization directly depends on propagation details, the properties of the medium ought to be known. For instance, uncertainties in the magnetic field could influence the polarization through Faraday rotation. 

At high energies ($E \gtrsim 1~\text{GeV}$) polarization is extremely hard to measure and virtually impossible in detectors based on electron-positron pair production (like \textit{Fermi} and \acp{IACT}). 
Recently, several satellite-based gamma-ray detectors were proposed, which use both Compton scattering and pair conversion detection techniques (e-ASTROGAM~\cite{DeAngelis:2016slk}, AMEGO~\cite{Kierans:2020otl}), enabling the measurement of polarization of gamma rays in the few MeV energy band. 
X-ray polarimetry has made a considerable step forward with the recent launch of Imaging X-ray Polarimetry Explorer (IXPE)~\cite{Soffitta_2021, 10.1117/1.JATIS.8.2.026002} and will likely improve even further with the upcoming eXTP~\cite{eXTP:2016rzs}.
Given the importance of gamma-ray polarimetry, there have been some suggestions on how to overcome these technical limitations~\cite{Ilie:2019yvs, Bernard:2022bnd}, in particular for the hard X-ray and soft gamma-ray bands~\cite{Bernard:2022jrj}.

The main requirement to improve the sensitivity of birefringence measurements would be to develop instruments capable of measuring polarization of high energetic photons.

\subsubsection{Effects from interaction anomalies}
\label{Sec:EffectsfromInteractionAnomalies} 

Some examples of analyses are reported in the following, for some specific effects and/or messengers, examining the sensitivity requirements to detect \ac{LIV} effects in point-source or diffuse fluxes, as discussed in \sref{Sec:InteractionAnomalies}. We note that we will emphasize the \ac{LIV} scenario in the discussion, as it is the one generally studied in the literature,  but one should not forget the possible alternative \ac{DSR} scenario with an energy scale much lower than the Planck scale.  In this alternative scenario one would have a modification of the fluxes of different high-energy astrophysical messengers when their energy approaches the energy scale of the deformation. The requirements to be able to identify this alternative scenario would be those allowing to exclude the correlation of effects in time delays and fluxes characterizing the \ac{LIV} case. Moreover, the energy dependence of the modification of the fluxes in the \ac{DSR} scenario will be different from the modification in the \ac{LIV} case in the energy range close to the energy scale of \ac{DSR}~\cite{Carmona:2021lxr}. Therefore, any improvement in the determination of the spectral flux of the different messengers would translate in an improvement in the sensitivity to effects of \ac{QG} in \iac{DSR} scenario.  

\paragraph{Effects in diffuse UHECR fluxes}

As mentioned in \sref{Sec:CosmicRaysInteractionAnomalies}, recent data suggests that cosmic rays at the highest energies are dominated by heavy nuclei, which limits their sensitivity to possible new physics effects, but the presence of a non-negligible subdominant component of protons cannot be excluded.
In order to improve the constraining power of new physics, the sensitivity to the proton fraction is one of the most relevant issues. The ongoing upgrades of existing \ac{UHECR} observatories and the new observatories planned for the next decades, listed in \sref{sec:proposals-UHECR}, should have a greatly improved discrimination power between protons and heavy nuclei, and collect a much larger amount of statistics than currently available. In addition, new analysis techniques such as the ones involving deep neural networks \cite{PierreAuger:2021nsq,PierreAuger:2021fkf} can be considered for the same aims. 

The requirement to observatories would be to determine the proton fraction in the region of the cutoff of the \ac{UHECR} spectrum, in order to support or discard the current interpretation of the \ac{UHECR} spectrum and mass data, and test the sensitivity to parameters of new physics. 
In parallel to the improvements of the \ac{UHECR} observatories, improvements in the modelling of hadronic interactions are expected to bring benefit to the interpretation of the \ac{UHECR} mass composition.

\paragraph{Effects in point-source or diffuse photon fluxes}

High-energy photons can be produced directly at the sources, resulting from leptonic processes, such as the self-synchrotron Compton mechanism, or from hadronic processes, as products of the neutral pion decay, or as subproducts of the interaction of \acp{UHECR} with the extragalactic backgrounds. These production processes, as well as the photon propagation during their path to Earth and in the atmosphere in the detection process, can be affected by \ac{LIV} physics. Although not many works have deepened in the consequences of \ac{LIV} in the production of photons at the sources (but see \cite{Tomar:2015fha} for related effects), mainly because of the complexity and uncertainties of source mechanisms, there have been some studies on \ac{LIV} limits from the development of atmospheric showers~\cite{Rubtsov:2016bea} and many others dwell on consequences from propagation effects~\cite{Galaverni:2008yj,Lang:2018yog,Terzic:2021rlx,Li:2022wxc}. 

Specifically, \iac{MDR} leads to a superluminal or subluminal velocity. Photon splitting and spontaneous emission put very strong constraints on a superluminal scenario from the detection of the highest energy gamma rays~\cite{LHAASO:2021opi,Li:2021tcw,Chen:2021hen}. They can be based on the lack of indication of a sharp spectrum cutoff when using data from identified luminous sources~\cite{LHAASO:2021opi,Chen:2021hen}, or based on the detection of extremely high-energy single events~\cite{Li:2021tcw,Chen:2021hen}, like the one with energy 1.4\,PeV detected by the \ac{LHAASO} collaboration~\cite{Cao:LHAASO} thanks to their very good rejection capability (the probability of this event to be a non-rejected cosmic ray is estimated to be 0.028\%~\cite{Cao:LHAASO}). To improve the sensitivity to \ac{LIV} effects, experiments will need to feature, besides excellent rejection properties, a good energy resolution (which for \ac{LHAASO} is $\sim 13\%$ at $100~\text{TeV}$ for showers with zenith angle less than $20^\circ$~\cite{Aharonian:2020iou}), and the ability to get a good estimate of the distance travelled by the gamma rays, which translates into a very good angular resolution (around 
$0.8^\circ$ at 10~TeV and $0.3^\circ$ at 
100~TeV for \ac{LHAASO}~\cite{Aharonian:2020iou}, which 
is however not good enough to firmly localize and identify the sources of the detected ultra-high-energy gamma rays~\cite{Cao:LHAASO}).

The \ac{LIV} effects are milder in a subluminal scenario, where there is a modulation of the energy spectra of gamma-ray sources, decreasing the absorption of gamma rays from their interaction with background photons. Constraints are usually derived in this case by comparing detected spectra with the propagation of a model for the intrinsic emission at the source in the LI and \ac{LIV} cases. The essential ingredients are therefore the selection of the spectra to be used in the analysis, the model of the intrinsic spectrum, and the choice of the model of background photons affecting the gamma-ray propagation~\cite{Lang:2018yog}. The best sensitivity to \ac{LIV} effects requires spectra characterized by a large distance to the source and the highest possible maximum measured energy, ingenious methods of obtaining the intrinsic spectra (beyond the standard approach of using the LI attenuation at the distance of each source, which is probably too naive for \ac{LIV} studies, as remarked in~\cite{Lang:2018yog}), and improved models for the \ac{EBL}, which still suffer from large uncertainties~\cite{Biteau:2015xpa,Dominguez:2010bv}.    

As for the source fluxes, the expected flux of cosmogenic photons in the subluminal scenario is affected by the \ac{EBL} modelling. In addition, cosmogenic photons depend strongly on the characteristics of the \acp{UHECR} emitted by their sources. Being mainly produced by the decay of neutral mesons, the amount of cosmogenic photons is connected to the \ac{UHECR} characteristics that maximize the photo-meson production; therefore, a contribution of protons at the highest energies at the escape from \ac{UHECR} sources would improve the ability to constrain \ac{LIV} effects in this sector, as shown in \cite{PierreAuger:2021tog}. An additional proton fraction at the sources that would increase the expected cosmogenic photon integral flux by 3 to 4 orders of magnitude at $10^{18}$~eV would allow reaching the sensitivity to cosmogenic photons at current UHECR observatories at these energies.

The current sensitivity to measure photons through the atmospheric cascade of particles they produce is obtained by taking into account the standard development of showers, as for instance in \cite{Niechciol:2023azz}. If \ac{LIV} effects are considered, as suggested in \cite{Rubtsov:2013wwa}, photons might escape observation passing through the atmosphere without producing air showers. The current limits should be therefore revised taking into account these possible effects.

\paragraph{Effects in diffuse neutrino fluxes}

The expected cosmic neutrino flux is also sensitive to parameters of new physics, which could modify the propagation of neutrinos in the extragalactic space due to non-standard neutrino oscillations and neutrino decays~\cite{Stecker:2014xja,Carmona:2022dtp}. These processes could affect the expected neutrinos produced directly in the sources, as well as the cosmogenic ones, namely those produced by the cosmic rays interacting with background photons while they travel in intergalactic space. In particular, changes in flavour composition and in the neutrino flux at Earth are expected, including the possibility of a cutoff in the spectrum  due to the fact that the decay probability strongly increases as the energy of the neutrino increases. The astrophysical neutrinos are sensitive to the scale of new physics responsible for changes of the flux in the energy region around 1~PeV \cite{Stecker:2014xja}. Therefore, we expect improvements from the increase of statistics of events in that energy region from IceCube or the next-generation neutrino experiments. The cosmogenic neutrino flux is expected to be originated in the interactions of cosmic-ray particles with the \ac{CMB} (contributing to the highest-energy peak of the flux) and with the \ac{EBL} (contributing to the lowest-energy peak). It is shown that increasing values of the parameters of new physics manifest themselves in re-shaping the \ac{EBL} or the \ac{CMB} peak~\cite{Reyes:2023osq}. 
For experiments trying to detect cosmogenic neutrinos, the enhancement of the exposure in the energy region where the \ac{EBL} peak is expected, between $10^{15}$~and $10^{17}$~eV, depending on the spectral index of the parent cosmic-ray flux at the escape from the sources \cite{Heinze:2015hhp,AlvesBatista:2018zui,Heinze:2019jou}, is of great relevance in order to improve the sensitivity to new-physics parameters. In fact, the expected number of neutrinos is doubled with respect to the \ac{SR} case, in the range $E_{\textrm{QG}}/M_\text{P}<10$ (considering a fixed scenario for the \ac{UHECR} spectral parameters and for the source evolution~\cite{Reyes:2023osq}).

\paragraph{Effects in neutrino oscillations}

We do not know whether the QG-motivated effects affect the 3 known flavours of neutrinos ($\nu_e$, $\nu_\mu$, $\nu_\tau$) differently, and one could expect they are the same for all flavours. However, astrophysical neutrino production models are asymmetric in flavours~\cite{Biehl:2016psj}. Even neutrino decays arising from a flavour-blind LIV can affect the flavour composition of astrophysical neutrinos, as indicated in \sref{anomalies-neutrinos}.  Besides this, LIV-driven neutrino oscillations have been intensively studied in an effective field theory approach~\cite{Kostelecky:2011gq}.

There are two key features we need to improve for future better sensitivity to these effects. 
First, the particle identification of neutrino flavour needs to be improved. Cherenkov detectors such as IceCube and Hyper-Kamiokande discriminate the events generated by muon neutrinos from the electron neutrino signals from morphology. Different detectors (for instance, the liquid scintillator of JUNO) will need to develop experiment-specific efforts such as the time profile of the signals. The distinction between events originated by a neutrino or antineutrino is also of great relevance to discriminate between effects originating from CPT-odd or CPT-even \ac{LIV} terms in the Lagrangian.

Second, energy and directional reconstruction needs to be improved for better resolution. The search for energy-dependent \ac{LIV} effects requires a good energy resolution which would make possible also a better binning separation of the data. Present and future experiments such as JUNO can take advantage of their unprecedented values of energy resolution. For the higher energy events, it will be important to perform calorimetric measurements and reconstruct the energies of through-going events.
With respect to angular resolution, this is relevant even in the case of isotropic \ac{LIV}, since even though its effect is energy dependent, the way it impacts oscillations of atmospheric neutrinos leads to a zenith-dependent effect through the dependence on the baseline~$L$ in the \ac{LIV} term, of the type~$L \, E^n$. For example, neutrinos travelling through the whole of Earth's diameter or neutrinos travelling horizontally to the detector will behave differently, due to this $L$~dependence. Moreover, the case of a sidereal-LIV is, by definition, direction-dependent. Last but not least, quantum decoherence can give also a zenith-dependent effect on neutrino oscillations.

\paragraph{Effects in the development of atmospheric showers} 
Astroparticles at the highest energies are investigated thanks to the measurements of the cascade of particles generated after the first interaction in the atmosphere. 
From this shower, one could aim to identify the potential sources. This would be possible with detectors with an angular resolution around or better than $1^\circ$ over all the energy range of interest. It is also important to have a huge area ground based detector based on a very robust detection technique which enables a duty cycle of about 100\% in case of transient sources.

Both the electromagnetic part of the shower and the muonic part can be affected by \ac{LIV} effects~\cite{Boncioli:2015cqa,PierreAuger:2021mve}. The change in the energy threshold of particle decays deplete the electromagnetic part of the shower faster than in the Lorentz invariant case, and the net effect is to move the shower maximum to higher altitudes, which consequently affects the calorimetric energy reconstruction of the shower.
With respect to the hadronic part of the shower, the decay versus interaction probability of pions can be altered in presence of \ac{LIV}, with some effects expected on the reconstruction of the energy and of the nuclear species of the primary cosmic ray \cite{Diaz:2016dpk,Klinkhamer:2017puj}. In particular, the relative fluctuations of the number of muons strongly decrease for protons as a function of the energy, if \ac{LIV} is included. Therefore, observables sensitive to the number of muons in the shower and to its fluctuations can be successfully exploited to test \ac{LIV} models \cite{PierreAuger:2021mve,Lobo:2021yem,Trimarelli:2023cmw}. 
Current bounds will benefit from the improved determination of the \ac{UHECR} mass composition as expected with the upgrade of the Pierre Auger Observatory. Other effects regarding the photon sector include the vacuum Cherenkov radiation in the atmosphere \cite{Duenkel:2023nlk} as well as the modified pair production explored in \cite{Rubtsov:2013wwa}, which could have an impact on the determination of the sensitivity for the observation of photons.

Currently, shower simulation codes such as CORSIKA~\cite{Heck:1998vt, CORSIKA:2023jyz} or CONEX~\cite{Pierog:2004re} do not allow the simulation of events that violate Lorentz invariance. It is therefore necessary to first model and then introduce modifications to these simulation codes to  compute the development of showers in the atmosphere in case of \ac{LIV}.

\medskip
In summary, the improvement of the sensitivity to LIV effects through interaction anomalies affecting the propagation of the cosmic messengers requires, quite generally, a better determination of their spectral fluxes, which will be achieved by addressing the following aspects:
\begin{itemize}
    \item very good energy and angular resolution;
    \item improved models for the \ac{EBL} and for the intrinsic spectra at the sources;
    \item the inclusion of LIV effects in atmospheric shower models;
    \item excellent rejection properties (e.g., between high-energy gamma rays and cosmic rays) and discrimination power (e.g., between protons and heavy nuclei, or between atmospheric and astrophysical neutrinos, and between neutrinos and antineutrinos);
    \item in general, much larger amount of statistics at high energies and the introduction of new procedures of analysis, such machine learning techniques. 
\end{itemize}

\section{Proposals for new measurement strategies}
\label{sec:proposals}

In this section, we discuss the ongoing and planned upgrades to existing observatories of cosmic messengers, proposals for future ones, and possible new approaches that could strongly benefit the research in \ac{QG} phenomenology. We divide this section into four parts in order to individually discuss the main cosmic messengers, namely gamma rays, neutrinos, \acp{UHECR}, and gravitational waves.

\subsection{Gamma-ray experiments}
The main targets of gamma-ray astronomy from the \ac{QG}-gravity perspective concern the possibility of in-vacuo dispersion and the possibility of \ac{QG}-corrected opacity of the universe to gamma-rays, as discussed in \sref{ssec:TD} and \sref{Sec:InteractionAnomalies}. Therefore, one desires to observe high-energy photons from very distant sources and those that can be described as short-duration bursts (or have intelligible fine time structure to enable time-of-emission considerations).
Gamma-ray telescopes with polarization measurement capabilities would have the added bonus of being able to investigate models in which the in-vacuo dispersion has a polarization dependence.

Clearly, the plans being made and implemented about the \ac{CTA} \cite{CTAConsortium:2017dvg}
are very exciting from the \ac{QG} perspective, since they will provide many opportunities for observing very-high-energy photons from fast and far transient sources. This in particular should include some observations of \acp{GRB} as well as observations of the Crab Nebula with its pulsar. One limitation of such observations is that it will deliver data just for relatively nearby sources, which not only decreases the expected magnitude of the effects (which could be compensated by the increase in the observed energies) but also lowers the sensitivity for investigating the form of the redshift dependence of possible in-vacuo dispersion. Complementing \ac{CTA} with telescopes capable of observing photons with energies between~1 and~100~GeV and sensitive to phenomena occurring at high redshifts would be very important for \ac{QG} research. The ideal option would be an upgraded version of the \textit{Fermi} \ac{LAT}.
An improvement of merely a factor~2 in effective area and sensitivity with respect to the currect \textit{Fermi} \ac{LAT} would already lead to progresses in \ac{QG} research.
Looking further in the future, a very desirable prospect would be the one of a network of a few of such ``super-\textit{Fermi}'' telescopes displaced at large distances (e.g.\ solar-system-scale distances).

High-altitude air-shower observatories are also a top priority for \ac{QG} phenomenology, as shown by the interest generated by results already reported by the \ac{LHAASO}~\cite{LHAASO:2023kyg}, located in the northern hemisphere. Besides that, \ac{QG} phenomenology and astrophysics in general would be benefited by a similar high-altitude instrument located in the southern hemisphere, such as SWGO~\cite{Albert:2019afb}.

\paragraph{Super-\textit{Fermi}} While all these strategies are extremely valuable, if one wants to imagine the ideal opportunity for \ac{QG} phenomenology this would be provided by the observation of the prompt phase of \acp{GRB} in the energy range between 10 and 100~GeV, particularly the prompt phase of short \acp{GRB}. This is challenging for the \ac{CTA} (moderate field of view, limited chances of seeing the prompt phase of \iac{GRB}) and for high-altitude air-shower observatories (good sensitivity only above 1~TeV), but could be achieved by planning a ``super-\textit{Fermi}'' space telescope, with larger effective area.

\paragraph{100+~GeV wide field of view ground-based observatories} If the spectral
break-off of gamma rays caused by internal absorption is larger than 100~GeV, a viable alternative/complement to
``super-\textit{Fermi}'' space telescopes could be provided by the SWGO \cite{Albert:2019afb} or the HADAR experiment \cite{Chen:2023inb}, planning a ground-based observatory with wide field of view and sensitivity to photons of energies 100~GeV and higher. Similar benefits for \ac{QG} phenomenology could come from the proposed  Plenoscope \cite{Mueller:2019cdz} and MACHETE \cite{2016APh....72...46C}.

\paragraph{All sky (multi satellite observatories)} Traditional single-satellite telescopes can cover the whole sky in a few hours, but their effective areas are not very large. 
There might be benefits for the ``discovery reach''
of \ac{QG} phenomenology
if there were space telescopes with much wider field of view (and therefore providing higher \ac{GRB} statistics), even if that came at the cost of a narrower and lower range of energies observed.
From this perspective \ac{QG} 
phenomenologists are following with strong interest the advent of ``distributed astronomy’’, using several nano-satellites that could serve as an all-sky monitor, with a keV--MeV energy band, providing high statistics at a small temporal scale, besides allowing a more accurate determination of the location of astrophysical events, like \acp{GRB}.  An example of this strategy is the GrailQuest mission \cite{HERMES-SP:2021mwp} (which would launch a fleet of hundred/thousands of nano-satellites by the 2050s), which is under development through the  HERMES project \cite{HERMES-SP:2021hvq}.

\paragraph{Laser-light experiments} Looking further ahead in the future, and assuming 
optimistically some rather large progress in our space-mission capabilities, one could hope for controlled experiments studying in-vacuo dispersion.
For example, if it will become possible to do laser-light experiments with lasers exchanging a signal between the Earth and the Moon, one could test in-vacuo dispersion using the frequency-doubler strategy proposed in \cite{Amelino-Camelia:2003egl}.

\subsection{Neutrino experiments}
Neutrino observatories 
are very relevant for in-vacuo dispersion studies \cite{Amelino-Camelia:2016fuh,Amelino-Camelia:2016ohi,Huang:2018ham,Huang:2019etr}. In-vacuo dispersion may affect particle oscillations and modify the flavour ratio at which the 3 known types of neutrinos ($e$, $\mu$, $\tau$, 
neutrino flavour) are detected on earth \cite{IceCube:2021tdn}. This effect could additionally be superposed by a possible family-dependence of in-vacuo dispersion (though none of the quantum-spacetime toy models that appeared in the literature supports this conjecture). As this effect grows with the energy of the particle and the distance from the source, astrophysical neutrinos play a significant role as they reach higher energies ($\sim \text{PeV}$) and have longer propagation distances ($\sim \text{Gpc}$).

The advent of a new generation of neutrino observatories, including 
KM3NeT~\cite{KM3Net:2016zxf} and IceCube-Gen2~\cite{IceCube-Gen2:2020qha}, will increase the statistics and improve the accuracy of energy determination, which is much more significant for the \ac{QG} research based on neutrino observations rather than for the astrophysics interest for neutrinos. Experiments based on other technologies such as air shower monitoring and radio arrays such as GRAND~\cite{GRAND:2018iaj}, can reach observations with higher energy~\cite{Arguelles:2022tki} and are therefore surely of great interest to \ac{QG} phenomenology. In planning these new observatories~\cite{Ackermann:2022rqc}, \ac{QG} research would be benefited by the accuracy of energy determination and increase in the statistics.

\subsection{Cosmic-ray experiments}\label{sec:proposals-UHECR}

As discussed in \sref{Sec:CosmicRaysInteractionAnomalies}, the main possible \ac{QG} effects on \ac{UHECR} observations are modifications of the kinematics of photonuclear interactions, which could alter the maximum distance from which \acp{UHECR} can be observed, and the production rates of cosmogenic neutrinos and photons. Also, modifications of particle lifetimes would alter the development of the air showers, as discussed in the final paragraphs of \sref{Sec:EffectsfromInteractionAnomalies}, and impact the unresolved ``muon puzzle'', which consists of a mismatch between the theoretically expected and the experimentally measured number of muons~\cite{PierreAuger:2021qsd,Albrecht:2021cxw}.

Presently our most powerful cosmic-ray telescope is the Pierre Auger Observatory, and its results have provided the basis for several \ac{QG}-phenomenology analyses \cite[and references therein]{Trimarelli:2023cmw}. 
However, in general converting \ac{UHECR} data into constraints on \ac{QG} parameters faces several challenges due to our lack of knowledge about the sources and propagation of \ac{UHECR}, described in \sref{sec:AstrophysicalModelling}, but one can be optimistic about progress in these directions in the coming years~\cite{Coleman:2022abf}.

As discussed in \sref{Sec:EffectsfromInteractionAnomalies}, concerning the planning of the next generation of cosmic-ray observatories, the top priorities for \ac{QG} are particle identification (especially distinguishing between cosmic-ray protons and cosmic-ray heavy ions), the detection of \ac{UHECR} photons and of course accuracy of energy determination.
From this perspective we are excited about AugerPrime \cite{Castellina:2019irv}, the ongoing upgrade of the Pierre Auger Observatory, adding scintillation surface detectors and radio detectors to the existing water Cherenkov detectors, which will greatly improve its sensitivity to the mass number of primary \ac{UHECR} nuclei. We are also looking forward to the ongoing upgrade of the Telescope Array (TA$\times$4 \cite{TelescopeArray:2021dri}), bringing its area from the present $700\,\text{km}^2$ to about $2800\,\text{km}^2$.
For the next decade we endorse enthusiastically the planning of GRAND~\cite{GRAND:2018iaj}, JEM-EUSO~\cite{Casolino:2023nld} and POEMMA~\cite{POEMMA:2020ykm}. As a successor of these facilities, the GCOS \cite{Horandel:2021prj, AlvesBatista:2023lqg} anticipates a set of arrays with total area of the order of $60\,000~\mathrm{km}^2$. For more details on the next generations of cosmic rays observatories, please refer to \cite{Coleman:2022abf}.
 
\subsection{Gravitational-wave experiments}
Our current and foreseeable capabilities of observation of gravitational waves do not look promising for tests of \ac{QG}-induced in-vacuo dispersion since these gravitational waves are of very long wavelength. The interest of the \ac{QG}-phenomenology community in gravitational-wave interferometry resides mainly in scenarios based on the idea of ``spacetime foam", such that \ac{QG} effects might manifest themselves as an additional source of noise \cite{Amelino-Camelia:1998mjq,Amelino-Camelia:2001dbf,Parikh:2020kfh}
for gravitational-wave interferometers.
While modelling of the conjectured \ac{QG} noise is still at a very preliminary stage, it appears \cite{Amelino-Camelia:1998mjq,Amelino-Camelia:2001dbf}
that the effects might be more noticeable at lower frequencies, which adds reasons of \ac{QG} interest in observatories like LISA \cite{LISA:2017pwj}, the
DECi-hertz Gravitational-wave Observatory (DECIGO) \cite{Yagi:2011wg}, 
NANOgrav \cite{NANOGrav:2023gor}, 
the Big Bang Observer (BBO) \cite{Yagi:2011wg,Mukhopadhyay:2021zbt} and the Einstein Telescope \cite{Maggiore:2019uih}. These planned (or ``planable'') gravitational-wave observatories might have added valence for \ac{QG} research through their ability to test some models of the 
stochastic background of gravitational waves 
\cite{Addazi:2018ctp,Addazi:2019dqt,Yang:2022ghc,Addazi:2022fbj}.

It has also been noticed that in some scenarios for spacetime quantization (see e.g.~\cite{Garcia-Chung:2020zyq, Garcia-Chung:2022pdy}), besides changes in their in-vacuo dispersion, gravitational waves may be affected with changes of their amplitude and frequency. Gradually as these studies reach greater maturity in the coming years, we would of course favour plans of gravitational-wave experiments tailored to these features.

From the multimessenger perspective it is interesting that gravitational-wave observations might be used in synergy with the in-vacuo dispersion studies based on data from gamma-ray telescopes. One can  hope that through some detections of gravitational waves we might obtain information suitable for improving \ac{GRB} models, which in turn could be valuable for studies of in-vacuo dispersion, for example, by providing insight into intrinsic spectral lags.

Finally, before closing the section, we should make the important remark, that as in the case of a cosmic refractive index (photons),  time delays of gravitational waves may be considered as indirect probes of dark matter, see e.g.\ \cite{Liao:2018ofi}. They can be sourced by the interference of the waves due to their scattering with primordial black holes of masses in the range $10$--$100M_\odot$ or planet-sized primordial black holes, which can play the role of a dark matter component.

\section{Data availability and collaboration between experiments}
\label{sec:data}
Before closing this white paper we would like to discuss future perspective of collaborations between researchers and collaborations working on \ac{QG} phenomenology and the openness of data. 

The community working on \ac{QG} phenomenology within multimessenger astrophysics is strongly interdisciplinary, combining theorists working on \ac{QG}, theorists working on astrophysics as well as experimentalists involved in different experiments, observing different types of messengers.
This perhaps also renders this community particularly sensitive to an incomplete transition in the policies that concern public availability of data. Until the early 2000s there were no open data requirements.
At the present time some experimental collaborations (mainly those which started operating several years ago) 
still work with limited (and in some cases absent) public availability of data, while others have adopted a strong commitment to publish and archive all data, documentation, tools and educational resources for using the data.
These differences probably also reflect corresponding differences in the policies adopted by funding agencies, some of which still pay no attention to public availability of data while others are placing an increasing emphasis on this point.
\par
Unfortunately in many cases a much larger fraction of the data collected by an experiment has been kept reserved for a much longer time period than would appear necessary for scientific reasons. For instance, the Pierre Auger Collaboration initially kept reserved 99\% of the \ac{UHECR} events detected (which was already an improvement over similar experiments, most of which did not publicly release any data at all until decommissioning), with the exception of the datasets used in a few specific works (most recently, all events with~$E \ge 32$~EeV detected until 31~December 2020 \cite{PierreAuger:2022axr}). In February 2021, it launched an Open Data Portal~\cite{auger-open}, but as of its writing~\cite{the_pierre_auger_collaboration_2022_6867688} it still excludes $90\%$~of the \ac{UHECR} events detected until 31~August 2018 and $100\%$~of the later ones. Nevertheless, there are plans to release up to~$30\%$ of its total data set, which exceeds the total exposure of all other \ac{UHECR} experiments combined~\cite{AbdulHalim:2023mD, PierreAuger:2023mbm}.
 
The situation is even worse with other similar experiments. For instance, the only list of \ac{UHECR} events released by the Telescope Array collaboration so far only includes events with~$E \ge 57$~EeV detected until 4~May 2013~\cite{TelescopeArray:2014tsd}.

It is important that funding of telescopes and observatories includes the necessary resources for a strong effort of public availability of data.
We do not propose a specific recipe, since the optimal solution strongly depends on the specifics of the proposed observations, and we well realize that it is also important to find a good balance between the idealistic interests of pure science and the practical interests of those devoting many years of their lives to preparing an observatory, who then use the embargo system for collecting deserved benefits for their efforts. 

But an aggressive data-release policy, requiring 100\%~of the data to be released within a fixed time period of detection (for instance of the order of the typical duration of a PhD program, e.g.\ five years), would impel collaborations to complete data analyses in a timely manner, and may end up being net beneficial for science. Moreover, the efforts to make the data usable to outsiders will also help keep the data accessible for archival studies long after the decommissioning of the detectors.

Also journal editors can have an important role, by being very forceful in requesting that data used in a scientific publication be made fully available to the  scientific community, also providing the information and tools necessary to understand and use the data.

We strongly endorse ongoing efforts by some collaborations (e.g.\ H.E.S.S. and MAGIC)
to improve their public-availability policies.
Two examples
which might be considered as reference standards are the \textit{Fermi} \ac{LAT} data policy and
the LIGO/Virgo data policy.
Both of them moved from an initial phase of full embargo of their data, to then making
an admirably strong commitment to release, archive, and serve the broader scientific community, also providing the information and the required tools to understand and use the data.
This combination of an initial phase of embargo followed by a phase of full disclosure might deserve to be adopted more broadly, and the benefits for science will be, of course, maximized if the embargo period is relatively short (the \textit{Fermi} \ac{LAT} had only one year of embargo, which could be an ideal choice, when other circumstances permit it).

While we are hoping that embargo phases will be short, when it happens that more than one telescope is within an embargo
phase (presently many telescopes are in full embargo) 
one practice that has tempered the impact on the progress of science is the establishment of working groups involving members of different experimental collaborations for the purpose of joint data analyses, governed by memorandums of understanding allowing them to access
each other's data before they are released to the general public.
Good examples of this practice are the Astrophysical Multimessenger Observatory Network (AMON), which uses subthreshold data (which is not suitable for astrophysical research) and public data from different observatories such as \ac{HAWC}, ANTARES, and IceCube to search for coincident multimessenger events \cite{AMONTeam:2022jnt,AMONTeam:2020otr}, the multimessenger observation of GW170817 \cite{LIGOScientific:2017ync}, the multimessenger observation of TXS 0506+056 \cite{IceCube:2018dnn}, the Working group on Hadronic Interaction and Shower Physics \cite{ArteagaVelazquez:2023fda}, ANTARES+IceCube+Auger+TA searches for neutrino-\ac{UHECR} correlations \cite{IceCube:2022osb}, Auger+TA anisotropy searches \cite{TelescopeArray:2014ahm} and the International Pulsar Timing Array \cite{Antoniadis:2022pcn}. These multimessenger initiatives can be boosted by platforms that emit alerts on astrophysical events involving several messengers, such as Astro-COLIBRI \cite{alert1}, General Coordinates Network (GCN) \cite{alert2}, Astronomer’s Telegram (ATEL) \cite{alert3} and the IceCube alert system \cite{alert4}. As for studies more directly related to searches for \ac{QG} effects, we point out the initiative \cite{Bolmont:2022yad} which has been involving researchers from H.E.S.S., MAGIC, VERITAS and LST-1 collaboration, and that describes a method to handle combined datasets from these telescopes in order to search for LIV-induced time delays.

\section{Final remarks}
\label{sec:final}

The realization that observations in high-energy astrophysics have the potential to reveal experimental signatures of a \ac{QG}
theory gave a significant boost to the field of \ac{QG} phenomenology since the 1990s.

Phenomena in extreme environments in the universe produce high-energy gamma rays, neutrinos, cosmic rays, and gravitational waves. Tiny \ac{QG} effects in the interactions and propagation of these messengers may show up in experimental observations thanks to the amplification offered by time delays accumulated in their travel through cosmological distances, or by threshold anomalies that may enhance or reduce significantly the expected detected spectrum at Earth.

The magnitude of these effects predicted by phenomenological \ac{QG} models depends on the modification they consider, a violation (\ac{LIV}) or a deformation (\ac{DSR}) of the symmetries of \ac{SR}, and on the basic parameter of the modification, the so-called \ac{QG} scale, $E_\text{QG}$. While this scale is usually associated with the Planck scale, it could also be considerably smaller, according to different theoretical scenarios.  \Ac{LIV} effects in the cosmic messengers are sensitive to a Planck-scale $E_\text{QG}$, but \ac{DSR} effects in the interactions are only relevant if $E_\text{QG}$ is close to the energies explored by these messengers, a situation whose consistency requires the absence of effects in the propagation in-vacuo.

The scale of $E_\text{QG}$ is therefore one of the most pressing theoretical unknowns, and experimental efforts in \ac{QG} phenomenology should aim for the identification of this scale. Unfortunately, given the current understanding of the consequences of a \ac{QG} theory, it is not possible to offer very specific proposals to achieve this objective. As a generic goal, this corresponds to trying to push the high-energy limit in the present and near-future experiments. This may be crucial for the observation of sharp cutoffs in the spectrum of photons or neutrinos predicted in superluminal \ac{LIV} models, as well as to observe an increase of the transparency of the universe for high-energy photons or cosmic rays in the case of \ac{DSR} or subluminal \ac{LIV} models. Another relevant effect that may be observed with better exposition at high energies is an enhancement in the cosmogenic production of neutrinos and photons by the interactions of cosmic rays with the \ac{CMB} and \ac{EBL}.

Besides the importance of trying to explore higher energies, improvements in the instrumental capabilities will directly reflect on the sensitivity to the \ac{QG} scale. 

In particular, better energy and angular resolutions are essential to establish correlations between neutrinos and photons that may be influenced by \ac{QG} effects. Also, these enhancements are of great importance in time delay studies, or to improve the localization of the sources of high-energy gamma rays, which is relevant for the study of interaction anomalies in their propagation. Excellent rejection properties, and, in general, an improvement in the sensitivity of the detectors to the very low fluxes of the most energetic messengers is also a key to disentangle \ac{QG} effects. Other upgrades that would be most welcome for this kind of phenomenology are more precise polarization measurements in the case of photons, better flavour and particle-antiparticle determination in the case of astrophysical neutrinos, an improvement in the discrimination power on the mass composition for the ultra-high energy cosmic rays, and better sensitivity to additional sources of noise for gravitational-wave interferometers. 

The previous experimental requirements need to be complemented with theoretical advancements in the modelling of the astrophysical processes that produce the cosmic messengers (relevant, for example, to discriminate between intrinsic and \ac{QG}-induced time delays), and of the backgrounds that affect their propagation, such as the \ac{EBL}, which are a source of systematic uncertainties in studies of the transparency of the universe. An improvement in the methods of analysis, trying to circumvent the limitations of the otherwise powerful likelihood method to model emission curves, or devising new strategies, as a proper combination of samples corresponding to sources of different redshifts, will also be important lines of future developments in the field. The inclusion of \ac{QG} effects in the simulation codes that model the propagation of the messengers or the development of showers in the atmosphere used for their detection is another important point here.

Progress along these lines will undoubtedly require comprehensive training for PhD students and young postdocs, equipping them with the necessary cross-disciplinary skills and techniques, both theoretical and experimental, essential for research in this field. The current white paper, along with the review on \ac{QG} phenomenology \cite{Addazi:2021xuf}, has resulted from the work conducted within \href{https://qg-mm.unizar.es/}{COST Action CA18108 ``Quantum Gravity Phenomenology in the multi-messenger approach''}. This COST Action also organized three Training Schools designed to provide students with the corresponding interdisciplinary expertise. The lecture notes from these schools are being published by PoS, and both the lecture notes and video recordings of the lectures are accessible on the webpage \url{https://qg-mm.unizar.es/publications/}. This resource will serve as a lasting legacy from the COST Action for the benefit of future generations of students.

Finally, we advocate supplementing the presently available set of observatories and their upgrades with a new generation of detectors, including new gamma-ray Earth-based and space instruments that allow for a better coverage of the sky and a new generation of neutrino and cosmic rays observatories that may be using new technologies such as radio arrays, with the purpose of providing higher statistics, better energy determination and particle identification, and an accurate location of astrophysical events. The complementarity between these detectors, sensitive to different energy ranges and to different messengers, makes \ac{QG} phenomenology strongly dependent on the cooperation and commitment of researchers working in different experimental collaborations, together with theorists working in the appropriate modelling of the new effects. Supporting the efforts of open data policies in these new facilities by funding agencies will therefore be essential for the progress of this highly interdisciplinary endeavour that may bring about deep changes in our way of thinking and describing the physical world.

\ack{The authors would like to acknowledge networking support by the COST Action CA18108 ``Quantum gravity phenomenology in the multi-messenger approach''.
RAB~is funded by ``la Caixa'' Foundation (ID~100010434) and the European Union's Horizon~2020 research and innovation program under the Marie Skłodowska-Curie grant agreement No~847648, fellowship code LCF/BQ/PI21/11830030\@. 
JMC\@, M.~Asorey, JLC~and MAR~are supported by the Spanish grants PGC2022-126078NB-C21, funded by MCIN/AEI/10.13039/501100011033, ERDF A way of making Europe; grant E21\_23R, funded by the Aragon Government and the European Union; and the European Union, NextGenerationEU Recovery and Resilience Program on `Astrofísica y Física de Altas Energías' CEFCA-CAPA-ITAINNOVA\@.
IL~was supported by the National Council for Scientific and Technological Development - CNPq Grant 312547/2023-4 and by the Grant 3197/2021, Para\'iba State Research Foundation (FAPESQ\@).
NEM~is supported in part by the UK Science and Technology Facilities research Council (STFC) under the research grants ST/T000759/1 and ST/X000753/1 and by the UK Engineering and Physical Sciences Research Council (EPSRC) via the research grant EP/V002821/1. 
CP~is funded by the excellence cluster QuantumFrontiers funded by the Deutsche Forschungsgemeinschaft (DFG, German Research Foundation) under Germany’s Excellence Strategy – EXC-2123 QuantumFrontiers – 390837967.
DRG~is funded the Spanish grants PID2022-138607NB-I00, funded by MCIN/AEI/10.13039/501100011033, ``ERDF A way of making Europe''.
HA~and GGL~would like to thank the Spanish ``Ministerio de Universidades'' for the awarded Maria Zambrano fellowship and funding received from the European Union - NextGenerationEU.
M.~Adamo, A.~Ballesteros, DFS\@, IGS\@, FM~and JJR~acknowledge support by the Q-CAYLE Proyect funded by the Regional Government of Castilla y Le\'on (Spain) and by the Ministry of Science and Innovation MICIN through NextGenerationEU (PRTR~C17.I1).
A.~Bonilla acknowledges a fellowship (44.291/2018-0) of the PCI Program - MCTI and CNPq.
The work of~MBL is supported by the Basque Foundation of Science Ikerbasque. Her work has been also financed by the Spanish project PID2020-114035GB-100 (MINECO/AEI/FEDER, UE\@). She would like to acknowledge the financial support from the Basque government Grant No. IT1628-22 (Spain).
ACO~and MG~gratefully acknowledge the financial support of the Spanish grant PID2019-107847RB-C42, funded by MCIN/AEI/~10.13039/501100011033.
SD~thanks the Natural Sciences and Engineering Research Council of Canada and the Alberta Government Major Innovations Fund for support.
MD\@, AÖ~and İS~acknowledge the support provided by the Scientific and Technological Research Council of Türkiye (TÜBİTAK\@).
JMD~acknowledges the support of project PID2022-138896NB-C51 (MCIU/AEI/MINECO/FEDER, UE) Ministerio de Ciencia, Investigación y Universidades.
GD~acknowledges the support provided by the Serbian Ministry for Science, Technological Development and Innovation under contract 451-03-47/2023-01/2000124.
AD~gratefully acknowledges the support from the European Union's Horizon research and innovation programme under the Marie Skłodowska-Curie grant agreement No.~101068013 with name QGRANT\@.
LD~is supported by a grant from the Transilvania Fellowship Program for Postdoctoral Research/Young Researchers (September 2022).
CER~acknowledges funding from PAPIIT UNAM Project TA100122.
SAFV~acknowledges support from Helmholtz-Zentrum Dresden-Rossendorf, PIP~11220200101426CO Consejo Nacional de Investigaciones Científicas y Técnicas (CONICET) and Project 11/X748 (UNLP\@).
LÁG~acknowledges the support of the Hungarian National Research Development and Innovation Offce (NKFIH) in the form of the Grant No.~123996.
TJ's research was supported by the Croatian Science Foundation Project No.~IP-2020-02-9614 Search for Quantum spacetime in Black Hole QNM spectrum and Gamma Ray Bursts.
JKG~and GR~acknowledge the support provided by the Polish National Science Center, project number 2019/33/B/ST2/00050.
JLS~acknowledges funding from ``The Malta Council for Science and Technology'' as part of the REP-2023-019 (CosmoLearn) Project.
CL~acknowledges partial support by a Boya Fellowship provided by Peking University and by the China Postdoctoral Science Foundation (Grant No. 2024M750046). His work was also financed by the Postdoctoral Fellowship Program of CPSF (Grant No. GZB20230032). 
BQM~acknowledges the support by National Natural Science Foundation of China~(Grant No.~12335006).
GM~acknowledges partial support by Conselho Nacional de Desenvolvimento Cient\'ifico e Tecnol\'ogico - CNPq under grant 317548/2021-2 and Fundac\~ao Carlos Chagas Filho de Amparo \`a Pesquisa do Estado do Rio de Janeiro - FAPERJ under grants E-26/202.725/2018 and E-26/201.142/2022.
DM~thanks the Julian Schwinger Foundation and the U.S.~Department of Energy (DE-SC00202062) for support.
VAM~acknowledges support by the Generalitat Valenciana via the Excellence Grant Prometeo CIPROM/2021/073, by the Spanish MICIN / AEI and the European Union / FEDER via the grant PID2021-122134NB-C21, and by the Ministry of Universities (Spain) via the mobility grant PRX22/00633.
SN~and JDZ~acknowledge grants PID2021-124591NB-C41,-C43 funded by MCIN/AEI/10.13039/501100011033 and by ``ERDF A way of making Europe''.
GJO~acknowledges support by the Spanish Grant PID2020-116567GB-C21 funded by MCIN/AEI/10.13039/501100011033 ``A way of making Europe'', and by the Generalitat Valenciana via the Excellence Grant PROMETEO/2020/079.
AÖ~also acknowledges the support received from the Abdus Salam International Centre for Theoretical Physics, Trieste, Italy.
RP~supported in part by the Swedish Research Council grant, contract number 2016-05996 and by the European Research Council (ERC) under the European Union's Horizon~2020 research and innovation programme (grant agreement No~668679).
VP~acknowledges the support by the Ministry of Education and Science of the Federation of Bosnia and Herzegovina, under project number 05-35-2467-1/23.
LP~acknowledges support by MUR (Ministero dell'Università e della Ricerca) via the project PRIN 2017 ``Taming complexity via QUantum Strategies: a Hybrid Integrated Photonic approach'' (QUSHIP) Id.~2017SRNBRK and financial support by the ``Angelo Della Riccia'' foundation.
A.~Platania acknowledges support by Perimeter Institute for Theoretical Physics. Research at Perimeter Institute is supported in part by the Government of Canada through the Department of Innovation, Science and Economic Development and by the Province of Ontario through the Ministry of Colleges and Universities.
SMMR~acknowledges the FCT grants UID-B-MAT/00212/2020 and UID-P-MAT/00212/2020 at CMA-UBI\@.
SR~acknowledges the support of the Natural Science and Engineering Research Council of Canada, funding reference No.~RGPIN-2021-03644 and No.~DGECR-2021-00302.
MAR~also acknowledges the FPI grant PRE2019-089024 by MICIU/AEI/FSE\@.
İS~also expresses gratitude for the support provided by the Anatolian University Libraries Consortium (ANKOS) and the Sponsoring Consortium for Open Access Publishing in Particle Physics (SCOAP3).
DS~acknowledges the support of Bulgarian National Science Fund No.~KP-06-N58/5.
AW~acknowledges financial support from MICINN (Spain) {\it Ayuda Juan de la Cierva - incorporaci\'on} 2020 No.~IJC2020-044751-I\@.
TT, JS, FR, JMC, GG and CP acknowledge the support from the University of Rijeka through project uniri-iskusni-prirod-23-24.
\\
{\bf{The following researchers endorse this white paper:}} \\
A.F.~Ali (Essex County College, NJ, USA);
M.~Arzano (University of Naples and INFN);
J-L.~Atteia (IRAP, Université de Toulouse, CNRS, CNES, UPS, Toulouse, France);
C.~Barceló (Instituto de Astrofísica de Andalucía, IAA CSIC);
I.~Ben-Dayan;
D.~Benisty (University of Cambridge, and Frankfurt Institute for Advanced Studies, fias);
N.~Bilic (Rudjer Boskovic Institute, Zagreb, Croatia);
O.~Birnholtz (Bar Ilan University, Israel);
D.~Blas ("Institut de Física d'Altes Energies (IFAE), BIST, 08193 Bellaterra (Barcelona), Spain and Institució Catalana de Recerca i Estudis Avançats (ICREA), Passeig Lluís Companys 23, 08010 Barcelona, Spain");
D.~Blixt (Scuola Superiore Meridionale, Italy)
D.E.~Bruschi (Institute for Quantum Computing Analytics (PGI-12), Forschungszentrum Jülich);
L.~Burderi (Department of Physics, University of Cagliari, Italy);
L.I.~Caramete (Institute of Space Science, Magurele, Romania);
M.~Chaichian (University of Helsinki, Finland);
M.~Chernyakova (Dublin City University);
S.~Clesse (Service de Physique Théorique, Université Libre de Bruxelles, Belgium);
M.~Demirci (Department of Physics, Karadeniz Technical University, Turkey);
P.K.~Dhankar (Department of Mathematics, G H Raisoni College of Engineering, Nagpur, India);
F.~Di Lodovico (King's College London);
T.~Di Salvo (Department of Physics and Chemistry, University of Palermo, Italy);
D.D.~Dimitrijević (Faculty of Sciences and Mathematics, University of Niš, Serbia);
J.~Ellis (Physics Department, King's College London, UK and Theoretical Physics Department, CERN, Switzerland);
C.~Escamilla-Rivera (Instituto de Ciencias Nucleares, Universidad Nacional Autónoma de México);
L.~Freidel (Perimeter Institute, Canada);
A.~Fuster (Department of Mathematics and Computer Science, Eindhoven University of Technology, The Netherlands);
J.~Gamboa (Universidad de Santiago de Chile);
A.~k.~Ganguly (Banaras Hindu University, India);
D.A.~Gomes (University of Tartu, Estonia);
M.J.~Guzman (University of Tartu, Estonia);
F.J.~Herranz (Departamento de Física, Universidad de Burgos, Spain);
A.~Ioannisian (ITPM and YerPhI, Yerevan 0036, Armenia);
P.~Jetzer (University of Zurich);
K.H.~Kampert (Department of Physics, University of Wuppertal, Germany);
A.~Karam (NICPB, Tallinn);
F.~Klinkhamer (Karlsruhe Institute of Technology);
G.~Lambiase (Università di Salerno, Italy);
C.~Lämmerzahl (ZARM, University of Bremen, Germany);
J.~Levi Said (University of Malta);
M.~D.~Maia, (Institute of Physics, University of Brasília, Brasília D.F., Brazil);
L.~Marchetti (University of New Brunswick, Canada);
P.~Martín-Moruno (Departamento de Física Teórica and IPARCOS, Universidad Complutense de Madrid, Spain);
H.~Martínez-Huerta (Departamento de Física y Matemáticas, Universidad de Monterrey, NL, México);
M.~Milošević (Faculty of Sciences and Mathematics, University of Niš, Serbia);
L.~Miramonti (University of Milano and INFN);
J.~Navarro-Salas (University of Valencia and IFIC, Spain);
S.~Navas (Dpto. de Fisica Teorica y del Cosmos and C.A.F.P.E., University of Granada, Spain);
M.~Niechciol (Experimentelle Astroteilchenphysik, Center for Particle Physics Siegen, University of Siegen, Germany);
E.R.~Nissimov (Insitute for Nuclear Research and Nuclear Energy, Bulgarian Academy of Sciences, Sofia, Bulgaria);
R.C.~Nunes (Instituto de Física, Universidade Federal do Rio Grande do Sul, Brazil);
N.A.~Obers (Niels Bohr Instistute and Nordita);
G.J.~Olmo (University of Valencia and IFIC, Spain);
A.~Ottewill (University College Dublin);
A.~Övgün (Physics Department, Eastern Mediterranean University, Turkey);
C.F.~Paganini (University of Regensburg, Germany);
U.~Pensec (Sorbonne Université, CNRS, France);
C.~Pérez de los Heros (Uppsala University);
B.~Puliçe (Sabancı University, İstanbul, Turkey);
A.~Racioppi (NICPB, Tallinn, Estonia);
M.D.~Rodríguez Frías (University of Alcala, Madrid);
H.~Sahlmann (Friedrich-Alexander-Universität Erlangen-Nürnberg, FAU);
İ.~Sakallı (Physics Department, Eastern Mediterranean University, Turkey);
M.~Sakellariadou (King's College London);
F.~Schussler (IRFU, CEA, Université Paris-Saclay, F-91191 Gif-sur-Yvette, France);
G.H.W.~Sigl (Universität Hamburg);
J.~Solà Peracaula (Universitat de Barcelona, UB);
C.F.~Sopuerta (Institute of Space Sciences, ICE, CSIC and IEEC);
N.~Stergioulas (Aristotle University of Thessaloniki);
T.S.~Stuttard (Niels Bohr Institute, Denmark);
T.~Thiemann (FAU Erlangen Nuremberg, Erlangen, Germany);
T.~Trześniewski (University of Wroclaw);
C.~Tzerefos (National and Kapodistrian University of Athens, Greece);
D.~Vernieri (University of Naples and INFN);
N.~Voicu (Transilvania University of Brasov, Romania);
A.~Wojnar (Department of Theoretical Physics and IPARCOS, Complutense University of Madrid, E-28040, Madrid, Spain);
D.~Zavrtanik (University of Nova Gorica, Slovenia) and
J.D.~Zornoza (IFIC, Univ. of Valencia - CSIC).
}

\section*{List of acronyms}
\begin{acronym}[IMAGINE]
    \acro{AGN}{active galactic nucleus}
    \acroplural{AGN}[AGNs]{active galactic nuclei}
    \acro{CMB}{cosmic microwave background}
    \acro{CTA}{Cherenkov Telescope Array}
    \acro{DSR}{doubly special relativity}
    \acro{EBL}{extragalactic background light}
    \acro{GR}{general relativity}
    \acro{GRB}{gamma-ray burst}
    \acro{HAWC}{the High-Altitude Water Cherenkov Observatory}
    \acro{IACT}{imaging atmospheric Cherenkov telescope}
    \acro{LAT}{Large Area Telescope}
    \acro{LHAASO}{Large High Altitude Air Shower Observatory}
    \acro{LIV}{Lorentz invariance violation}
    \acro{MDR}{modified dispersion relation}
    \acro{SR}{special relativity}
    \acro{SWGO}{Southern Wide-field Gamma-ray Observatory}
    \acro{UHECR}{ultra-high-energy cosmic ray}
    \acro{QG}{quantum gravity}
\end{acronym}

\providecommand{\aap}{Astron.\ Astrophys.}
\bibliographystyle{iopart-num}
\bibliography{CA18108bib}
\end{document}